\documentclass[letterpaper]{article}

\usepackage{cureport}

\usepackage{algorithm}
\usepackage{algpseudocode}
\usepackage{graphicx}
\usepackage{subfigure}
\usepackage{epsfig}
\usepackage{url}
\usepackage{listings}

\TechReportYear{2011}
\TechReportMonth{November}
\TechReportNumber{1085-11}

\begin{document}

\title{Food Redistribution as Optimization}
\author{Caleb Phillips, Rhonda Hoenigman, Becky Higbee, and Tom Reed}
\maketitle

\begin{abstract}
In this paper we study the simultaneous problems of food waste and hunger in the context of the possible solution
of food (waste) rescue and redistribution. To this end, we develop an empirical model that can be used in Monte Carlo simulations
to study the dynamics of the underlying problem. Our model's parameters are derived from a unique data set provided
by a large food bank and food rescue organization in north central Colorado. We find that food supply is a non-parametric
heavy-tailed process that is well-modeled with an extreme value peaks-over-threshold model. Although the underlying process
is stochastic, the basic approach of food rescue and redistribution appears to be feasible both at small and large scales. The ultimate efficacy of 
this approach is intimately tied to the rate at which food expires and hence the ability to preserve and quickly 
transport and redistribute food. The cost of the redistribution is tied to the number and density of participating 
suppliers, and costs can be reduced (and supply increased) simply by recruiting additional donors to participate. 
Our results show that with sufficient funding and manpower, a significant amount of food can be rescued from 
the waste stream and used to feed the hungry.
\end{abstract}

\section{Introduction}

There is a contradiction present in the United States (US) today:
between 27\% and 50\% of food produced for consumption is wasted in some stage of production, distribution, or preparation \cite{Hall2009,Jones2005,Kantor1997}.  
Meanwhile 14.7\% of americans (1 in 7) are having difficulty finding enough to eat,
and 5.7\% are going hungry\footnotemark. The populations most at-risk of hunger include 36.6\% of households with children
and single mothers, 27.8\% of households with children and single fathers, 26.9\% of
hispanic households, 24.9\% of black households, 21.3\% of households with children, and 7.8\% of seniors living alone \cite{Nord2009}. Globally, food prices have reached unprecedented levels in 2011, and we are
currently in the midst of a worldwide hunger epidemic \cite{economist}. A clear question arises when studying these statistics: is it possible
to recover food from the waste stream and redistribute it to those who are hungry, thereby reducing both waste
and hunger?  

\footnotetext{The United States Department of Agriculture (USDA) defines these two classes of hunger as ``low food security'' and ``very low food security''.}

The idea of food rescue and redistribution is not a new one. Non-profit food rescue and gleaning organizations 
(e.g., \cite{cityharvest,philabundance,seniorgleaners}) have been operating on this basic premise for
more than 30 years, and there are dozens of organizations of varying size that currently rescue food in some capacity and redistribute it. These organizations recover food that would otherwise be wasted from \textit{donors} (e.g., grocery stores, farms, retailers, restaurants) and redestribute it via \textit{agencies} (e.g. food banks/pantries, soup kitchens, and shelters) to those in need. Recently, a coalition of major grocers and retailers organized under the Feeding America project with the goal of large scale food rescue, redistribution, and documentation \cite{FeedingAmericaReport}. Two popular recent books have studied the problem of systematic food waste in both the U.S. and Europe \cite{Bloom2010,Stuart2009}. Yet, to our knowledge there has been no prior effort to quantify the cost and practicality of the food rescue and redistribution model on a large scale.

Identifying sustainable systems for reducing food waste and hunger and understanding their cost, can have a substantial impact on waste reduction and food distribution on a local, national and even global scale. In this paper, we make an important first step in this direction by studying the food rescue and redistribution dynamics in a pair of neighboring counties in north central Colorado. We investigate the task of food recovery as a time-sensitive spatial distribution problem involving food supply and demand and the energy cost of redistribution from donors to agencies. To this end, we build an empirical model using data from a large food rescue organization in north central Colorado. Although our results are limited to this region, we contend that it is representative of a large class of similar regions with a mix of rural, small urban, and urban service areas. Using this model, we present an optimization framework for finding low cost solutions to food recovery and redistribution, which can be used to determine average, best case, and worst case bounds on the problem through simulation.

The main contributions of this work are as follows:

\begin{itemize}
\item We present the first formal description of the food redistribution problem as a well defined optimization problem.
\item We investigate a novel data set of food rescue and distribution from a large food bank and discover that the food supply process is heavy tailed and can be well modeled with extreme value theory.
\item We present initial results on the feasibility of the redistribution approach and find that basic demand can be easily met with available supply, and that with small changes and optimizations, a substantial portion of hunger may be mitigated with rescued food.
\end{itemize}

\begin{figure}
\begin{centering}
\includegraphics[width=0.6\columnwidth]{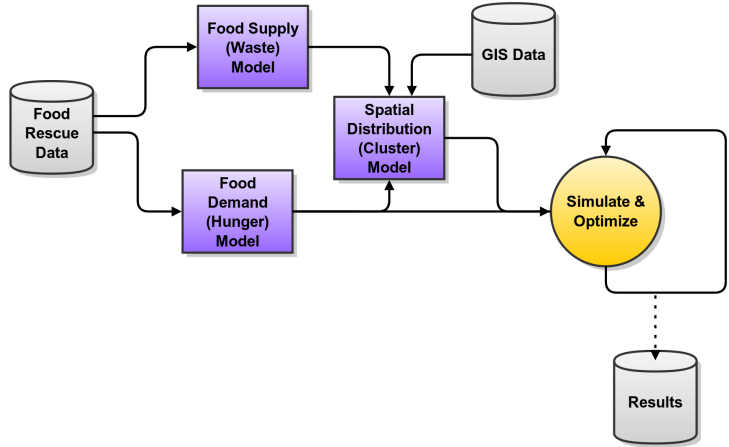}
\caption{Overview of modeling and simulation process.\label{fig:schema}}
\end{centering}
\end{figure}

Figure \ref{fig:schema} provides an overview of the modeling and simulation process used in this paper. First, using food rescue data, we build statistical models for food supply (waste) and food demand (hunger). Next, Geographical Information Systems (GIS) data is used to locate and cluster nearby donors. We also measure the square footage of each donor at this point to determine its size, and compute driving routes between all pairs of donors. Finally, the cluster model, demand model, and supply models are used in a Monte Carlo style stochastic simulation. At each iteration (day), an optimal schedule is determined that both meets demand (if possible) and minimizes cost (kilometers driven). The output of the simulation is then studied to assess feasibility and quantify costs.

In the next section we describe the data collected for this study and the statistical models we derive from it. In section \ref{sec:model}, we discuss the other features of our model, including the optimization framework and how it fits in with similar problems. Section \ref{sec:implementation} describes the implementation of our simulation framework and design decisions. Section \ref{sec:experiments} discusses our experimental design and section \ref{sec:results} provides the results of the experiments. Finally, section \ref{sec:fin} concludes with a summary and discussion of future work.

\section{Measurements}
\label{sec:data}

The data we use in this study was supplied by Community Food Share (CFS), the sole food bank for Boulder
and Broomfield counties in north central Colorado \cite{CFS}. This data includes the pounds of food received from each donor on each day for from July 1, 2010 to August 31, 2011. There were donations from 90 distinct donors, comprising 20,270 donations and 2,328,821 lbs of food\footnote{Most food rescue organizations weigh their donations so that they can provide documentation to their donors for the
purpose of tax-deduction. CFS uses categories based on USDA food ``pyramid'' groups: grains, vegetables, fruit, protein, dairy, fat/sweets, 
soup/baby food, and non-food.}\footnote{Donations are broken up by food category, so an individual pickup may be documented with as many as 7 distinct donations. Some donations are recorded by case and are not weighed. In this case we have taken a lower-bound estimate that one case weighs 25 pounds. Although the weight of a case varies by product, we have used 25 because CFS suggested this was a reasonable average estimate.}. This food was distributed to 304 unique agencies, which are predominantly homeless shelters, soup kitchens, smaller food pantries, and other organizations that serve at-risk populations. 

\begin{figure}
\begin{centering}
\subfigure[By Week]{\includegraphics[width=0.8\columnwidth]{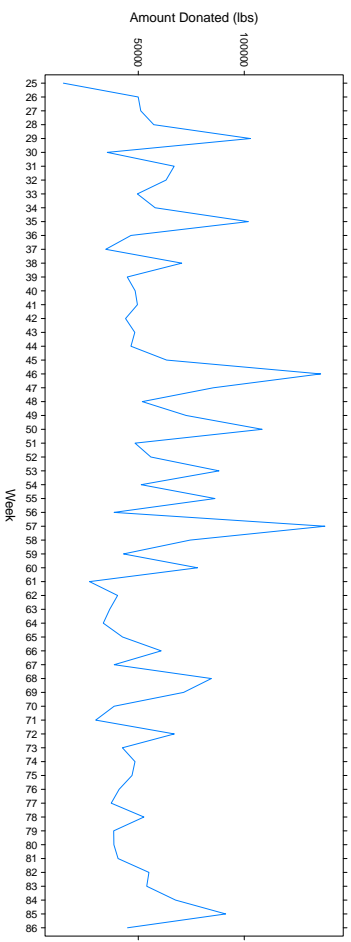}}
\subfigure[By Month]{\includegraphics[width=0.8\columnwidth]{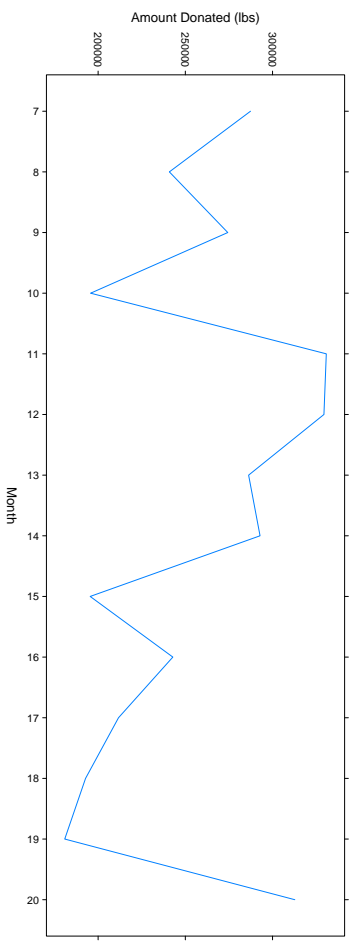}}
\subfigure[By Day of Week]{\includegraphics[width=0.8\columnwidth]{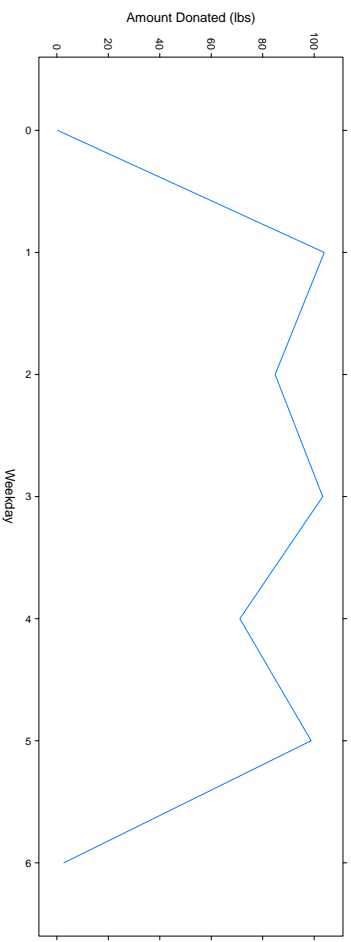}}
\caption{Temporal variation in amount of donations received by CFS.\label{fig:temporal}}
\end{centering}
\end{figure}

Figure \ref{fig:temporal} provides an overview of this data, showing the amount of food received each week and month, as well as the average amount received each week day. We can see some clear annual trends in this data. There is a trend upwards in the winter months when CFS has the majority of its food drives. \footnote{Although we have excluded food drives in our data, the increased exposure they create may cause an increase of food donations around the same time.} The daily donations are approximately the same amount of food, with the exception of the weekend, when CFS is closed. As a result, donations on Monday and Friday appear to be slightly higher than the days in the middle of the week.

\begin{figure}
\begin{centering}
\includegraphics[width=0.45\columnwidth]{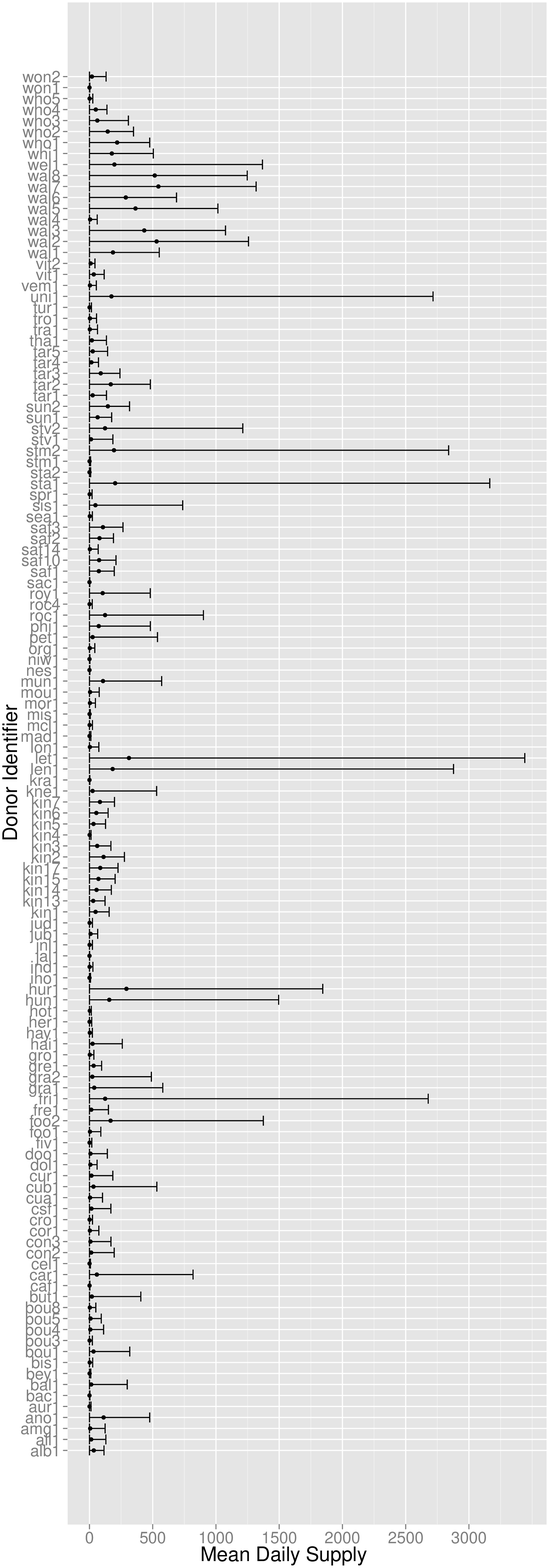}
\caption{Mean Daily supply data by donor in pounds.\label{fig:supplyhist}}
\end{centering}
\end{figure}

\begin{figure}
\begin{centering}
\subfigure[Grocers]{\includegraphics[width=0.4\columnwidth]{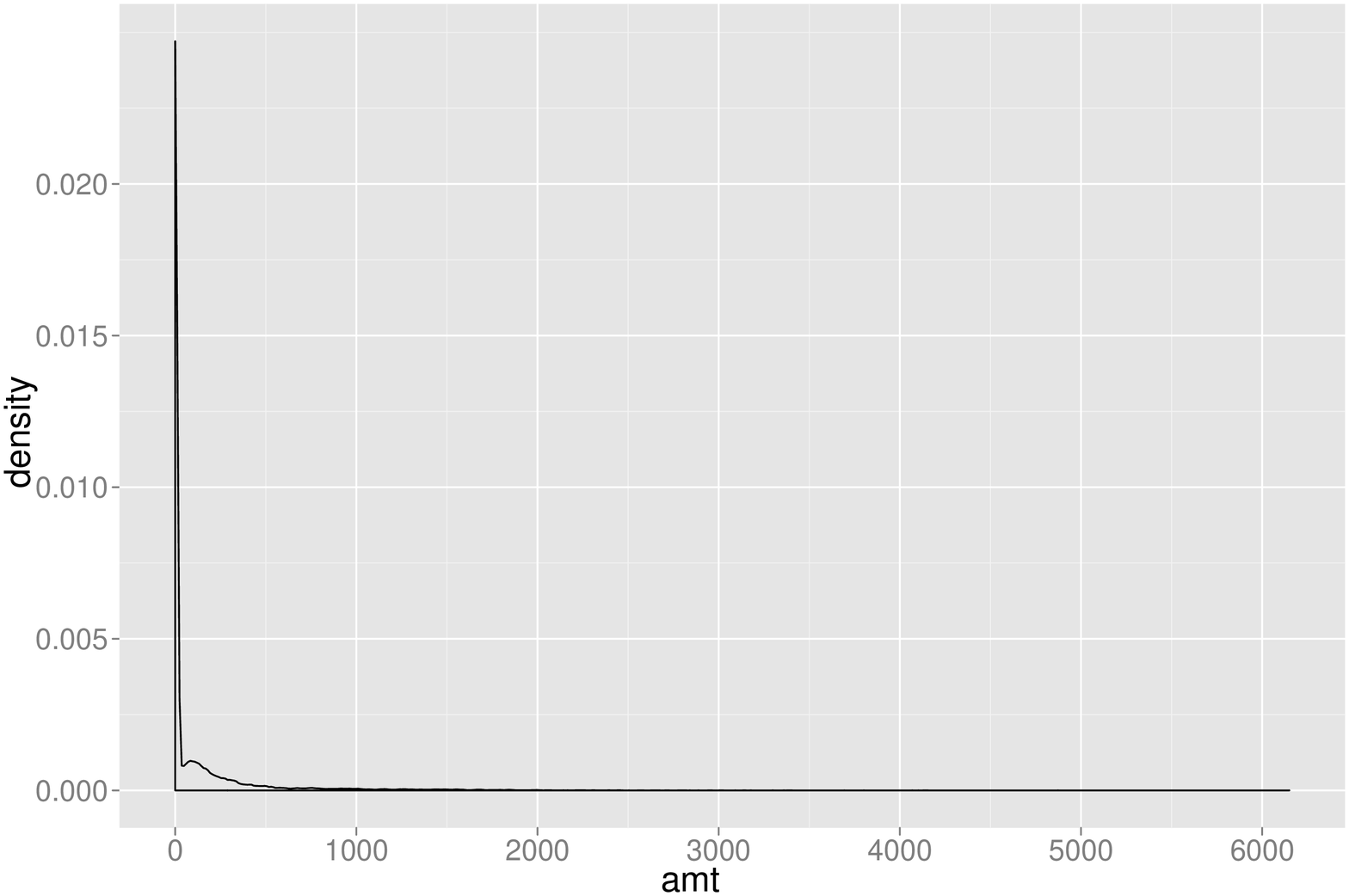}}
\subfigure[Manufacturers]{\includegraphics[width=0.4\columnwidth]{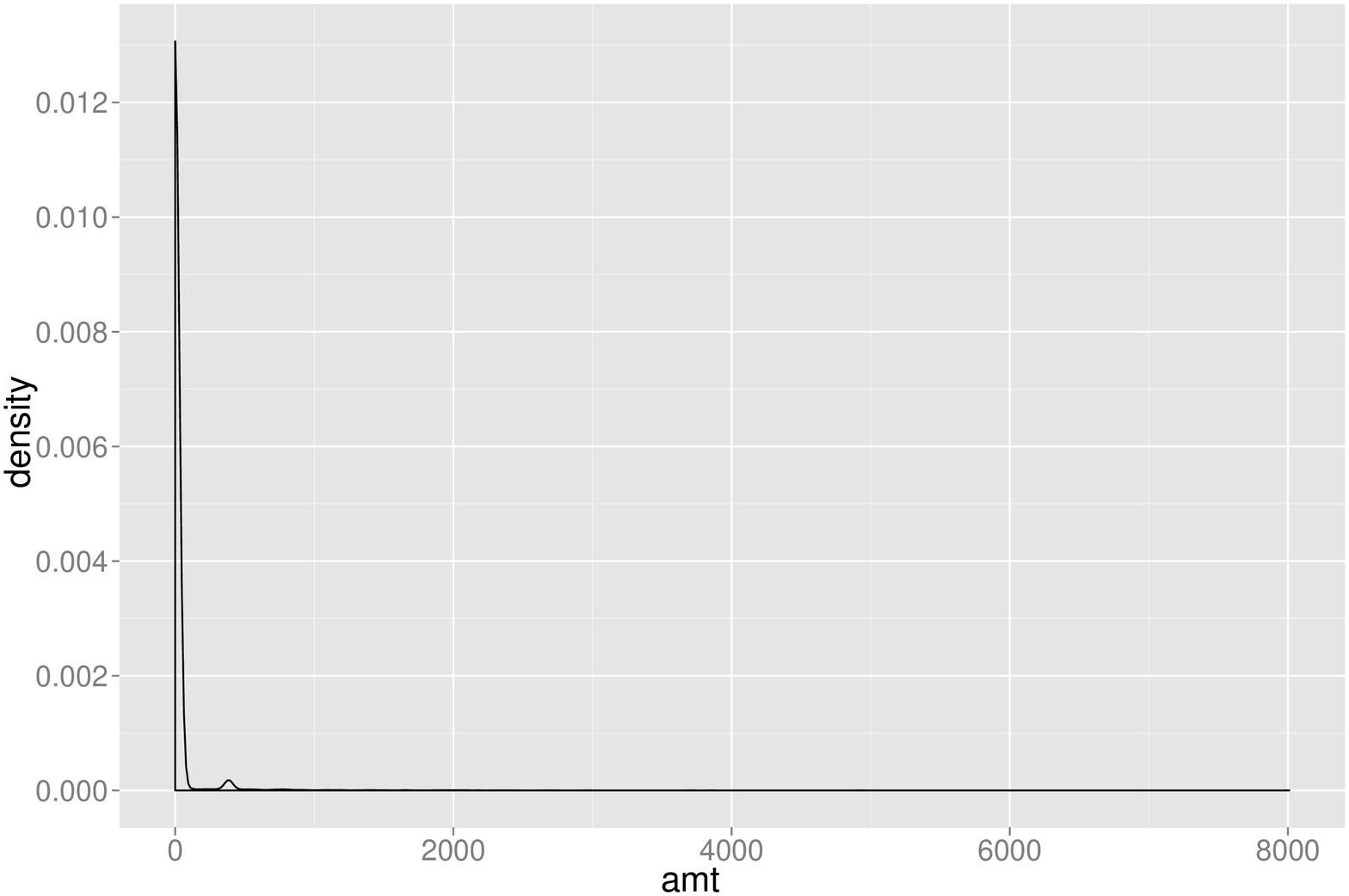}}

\subfigure[Farms]{\includegraphics[width=0.4\columnwidth]{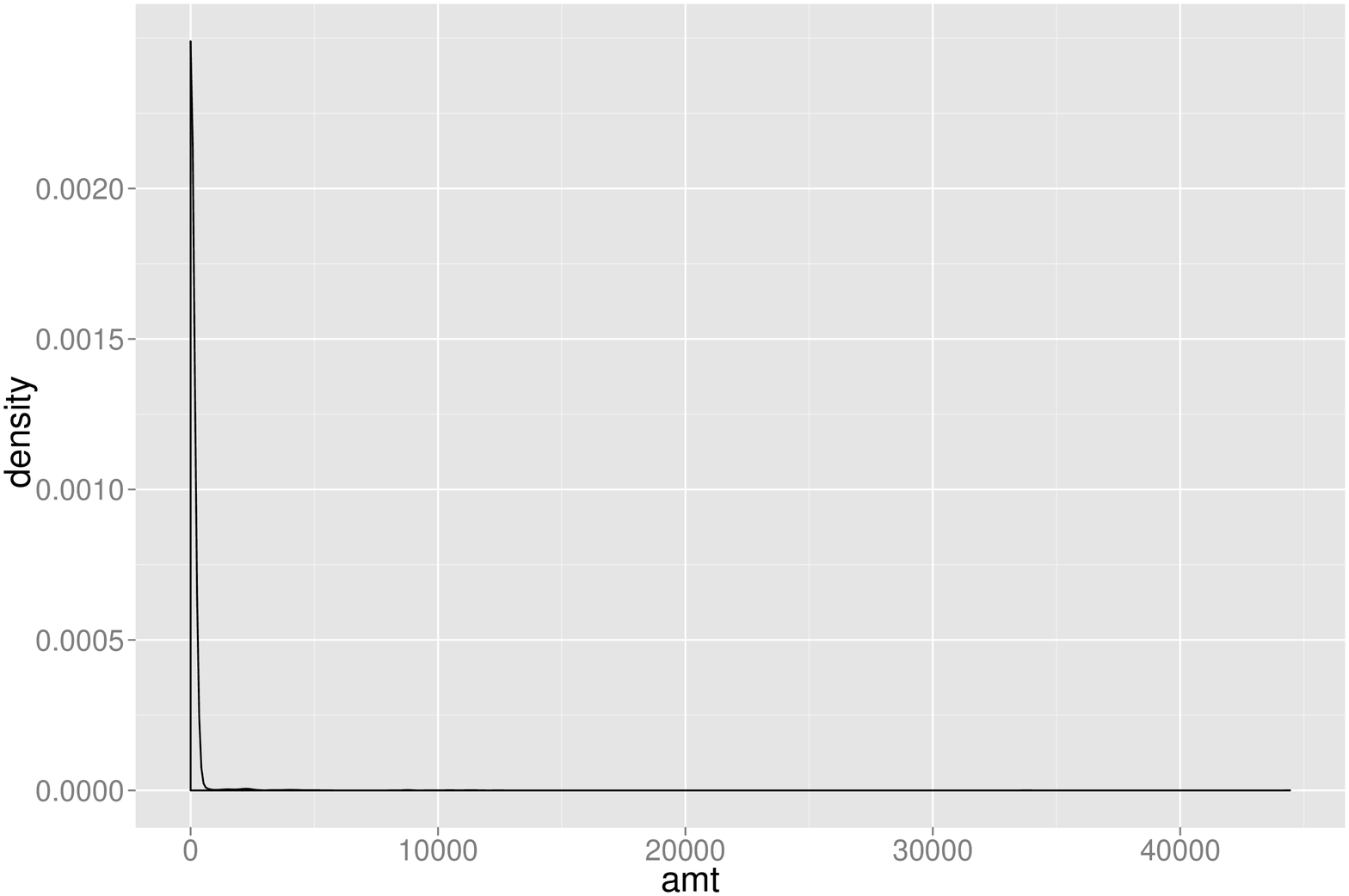}}
\subfigure[Individuals]{\includegraphics[width=0.4\columnwidth]{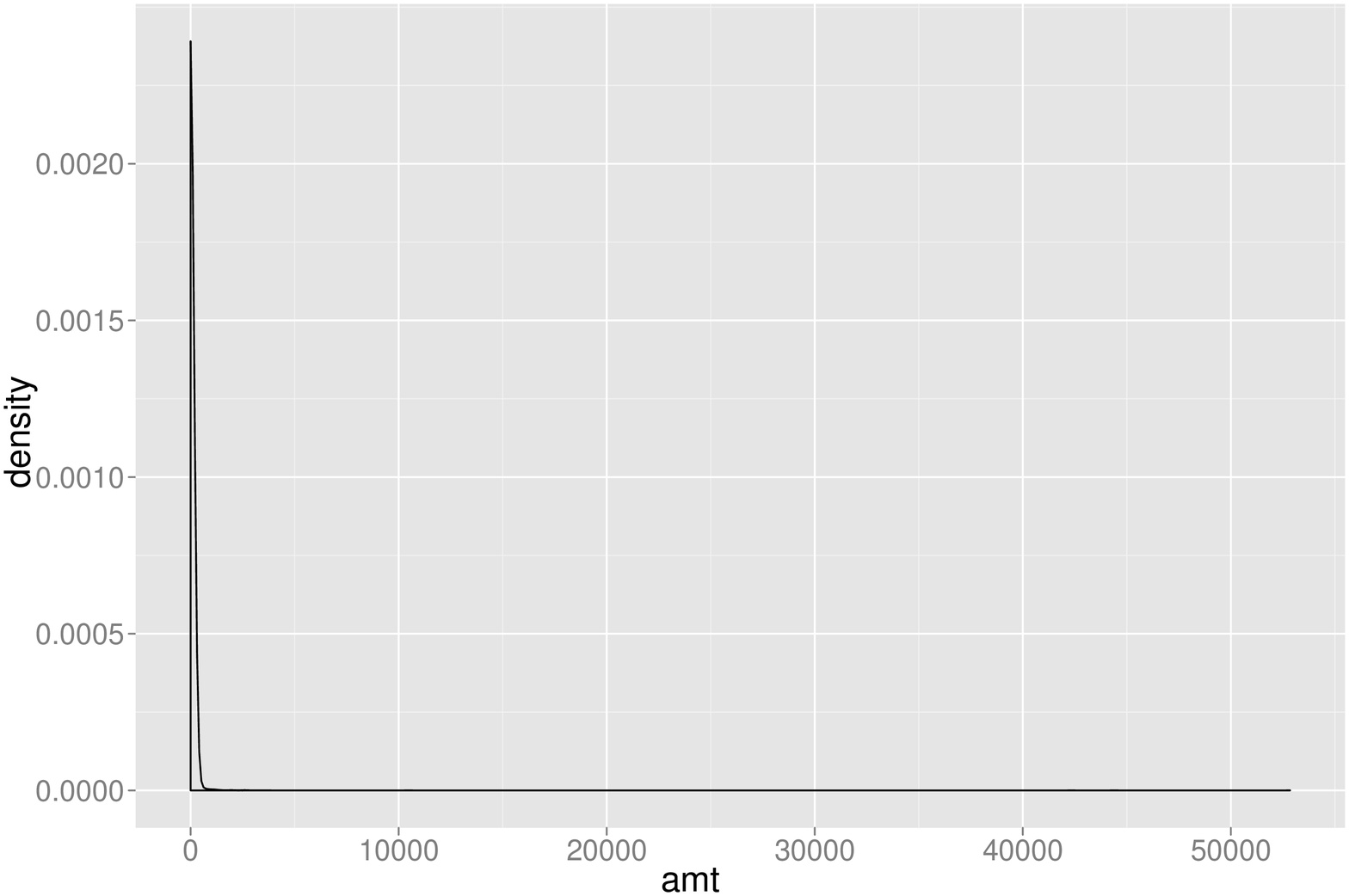}}
\caption{Distribution of supply by category. The individuals category contains donations from both (not food related) corporations and citizens.\label{fig:supplypdf}}
\end{centering}
\end{figure}

\begin{figure}
\begin{centering}
\subfigure[PDF]{\includegraphics[width=0.49\columnwidth]{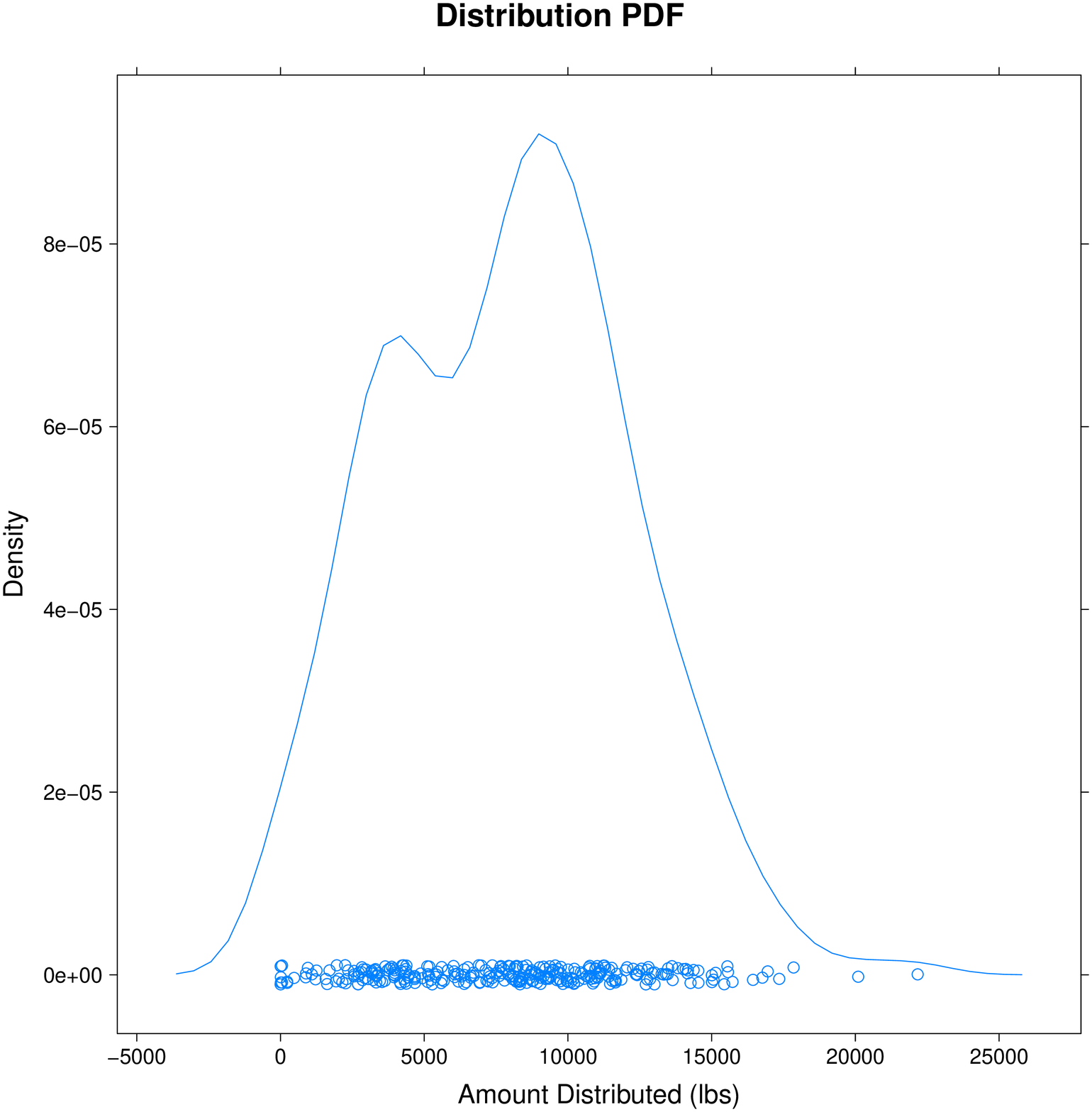}}
\subfigure[Normal QQ-Plot]{\includegraphics[width=0.49\columnwidth]{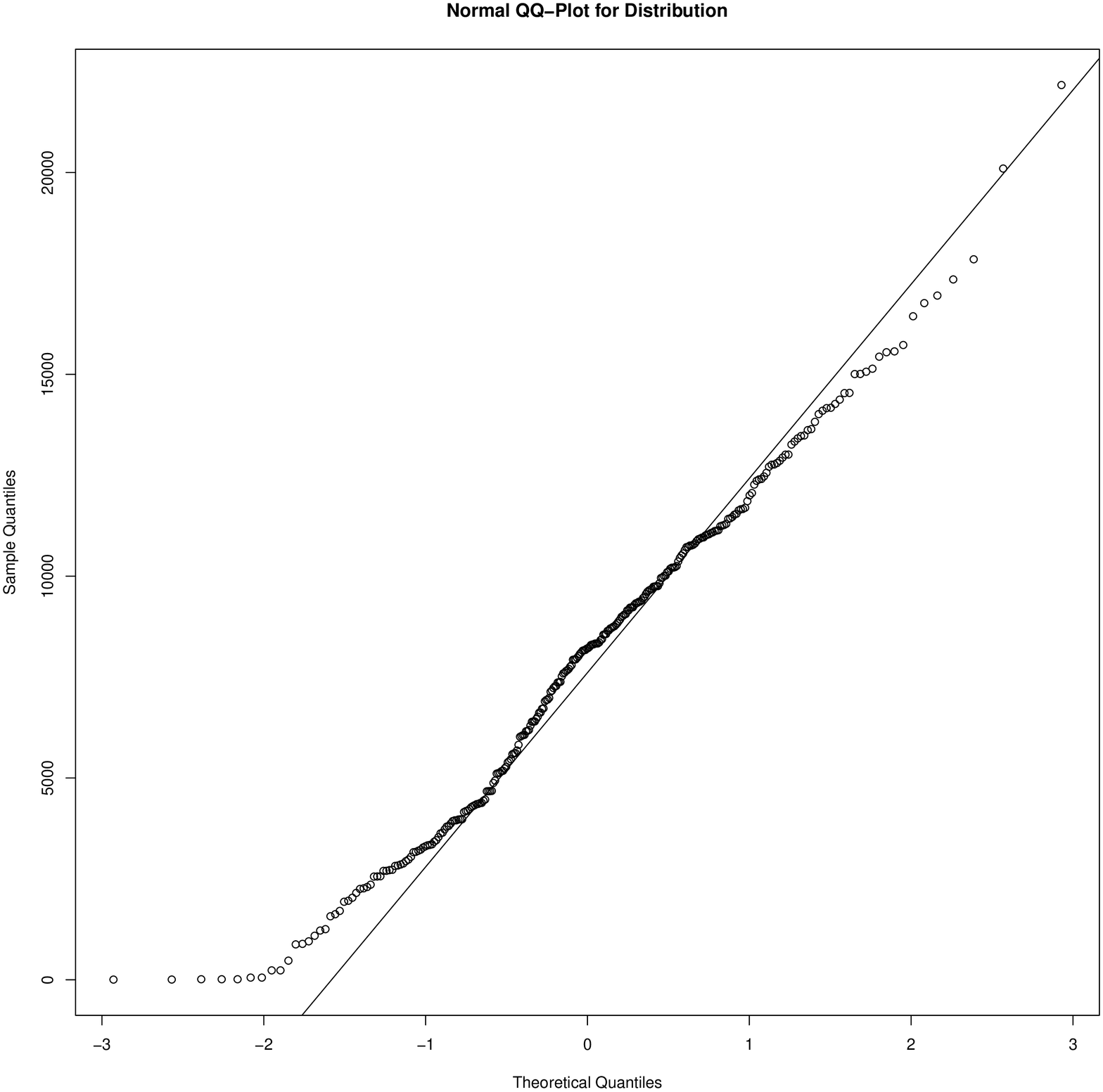}}
\caption{Daily total receiving data in pounds.\label{fig:supp:pdf}}
\end{centering}
\end{figure}

Figure \ref{fig:supplyhist} plots the mean daily food supply (in lbs) from each donor and figure \ref{fig:supp:pdf} shows the aggregate distribution of donated food, which appears to be Gaussian with a mean of 5,454 and standard deviation of 5,051. Although we can use this distribution directly to model the aggregate supply from all donors, we must characterize the per-donor supply to enable simulation. To this end, we subdivide donors by category; figure \ref{fig:supplypdf} plots a probability distribution function (PDF) of the daily donations where the donors are divided into: grocers, manufacturers, farms, and individuals. The individuals category contains donations both from named individuals and anonymous donors, as well as donations from corporations and organizations that are not in the food-service industry. We have purposely excluded donations from sister food banks and food drives, which are generally composed of purchased rather than rescued food. Some donors provide substantially more food than others and the largest grocers provide the most food on average. We will look at correlations between store size and amount donated in section \ref{sec:predsupp}. The PDF hints that the underlying data is heavy-tailed, with zero-supply as a normal case and large supply events occurring with smaller probability. Although each of the categories has its own shape and scale, the same basic heavy-tailed distributional shape seems to dominate.

\subsection{Characterizing Supply (Waste)}
\label{sec:data:supply}


We use the data provided by CFS to statistically charactize the dynamics of the supply and demand processes. On a given day, the data includes some number of donations, or ``food supply events.'' For a given donor, we make the assumption that on days where food is not picked up, there is zero supply. This data has a distinct heavy tail, where distribution is skewed to the left, indicating that small values are most common, but that with small probability, large and sometimes extremely large values can be observed, as seen in figure \ref{fig:supplypdf}.

This type of distribution is well modeled by a Peaks Over Threshold (POT) approach, using a fit from a Generalized Pareto Distribution (GPD), which is a power law probability distribution that features a long heavy tail, i.e., most events are small, but some events can be extremely large. The POT approach to fitting is common in weather modeling, and particularly in modeling the likelihood of extreme weather events \cite{Coles2001}. In this approach, there is a threshold value ($\theta$) for an event that is selected at some constant rate ($r$). In our problem the threshold value is zero, and we are looking to model the probability and size of food supply events greater than that threshold---where the donor has waste that can be recovered. Events larger than the threshold are modeled with a heavy tailed distribution described by location ($\mu$), scale ($\sigma$), and shape ($\zeta$) parameters. These parameters describe the basic shape, width, and location shift of our fit. The location parameter, which is the mean in the parametric sense (i.e., it is the ``center'' of the entire distribution), is zero. The scale parameter describes the uncorrected central tendency of the tail and the shape parameter describes its spread.

\begin{figure}
\begin{centering}
\includegraphics[width=0.8\columnwidth]{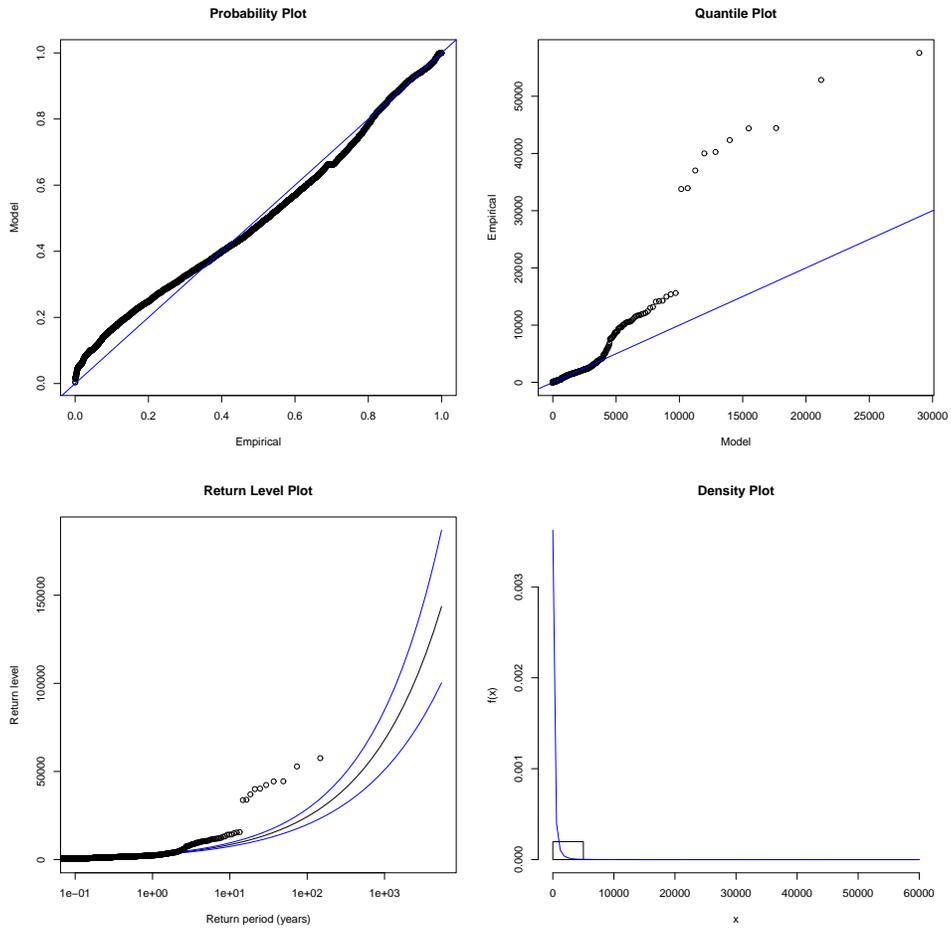}
\caption{Generalized Pareto fit statistics for all suppliers' data.\label{fig:gpd:supply:1}}
\end{centering}
\end{figure}

\begin{figure}
\begin{centering}
\includegraphics[width=0.8\columnwidth]{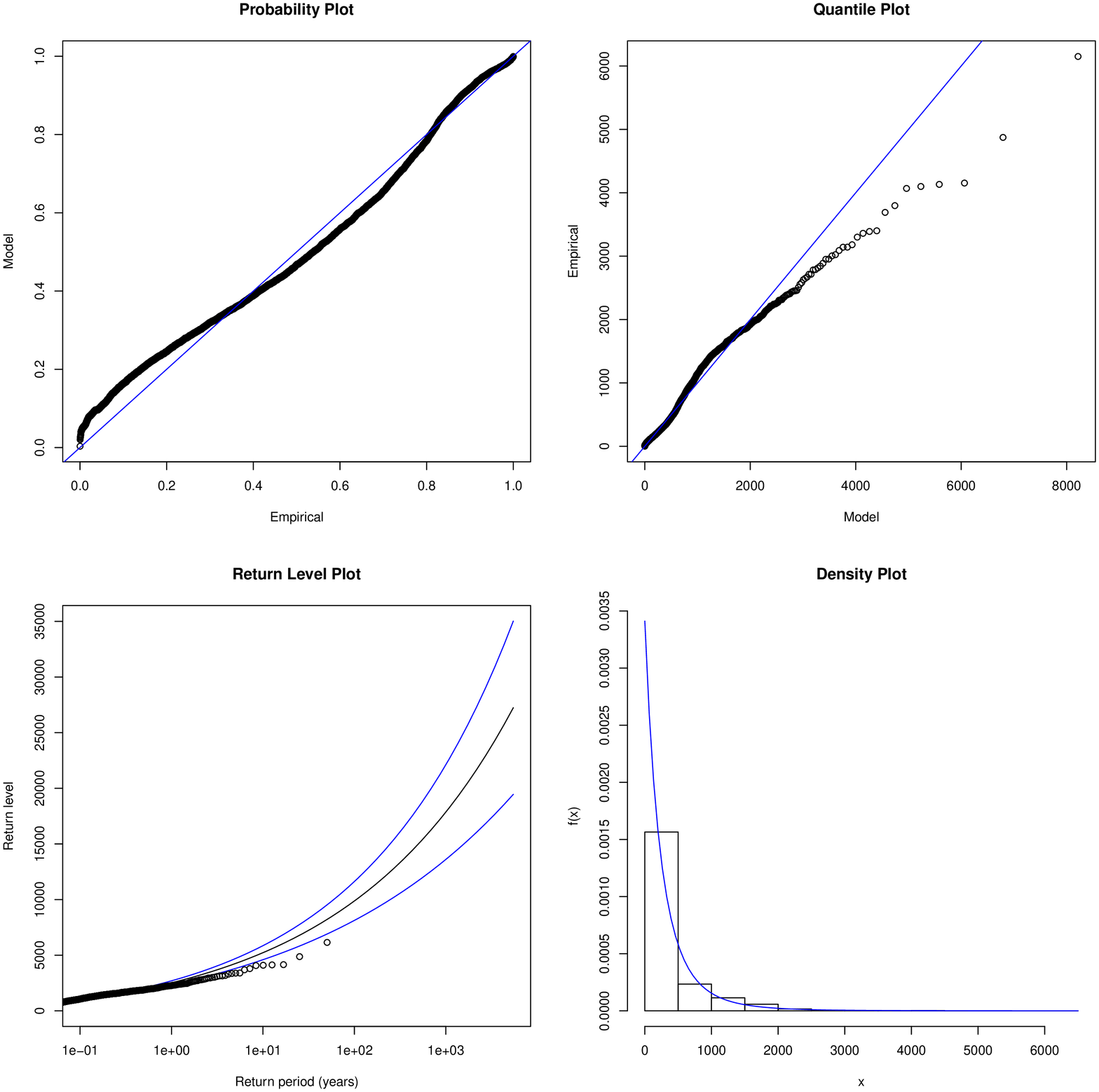}
\caption{Generalized Pareto fit statistics for grocers' supply data.\label{fig:gpd:supply:2}}
\end{centering}
\end{figure}

In order to acheive tighter fits, we model the supply data for all donors as well as manufacturers, grocers, individuals, and farms separately. Figure \ref{fig:gpd:supply:1} shows the quality of the fit for all donors combined, and figure \ref{fig:gpd:supply:2} shows the fit for grocers alone. These figures provide the probability plot, quantile-quantile plot, and return level plot, which visually compare the distribution of the observations and the fitted model. In these plots, the closer the points are to the diagonal line, the better the model is able to predict the data. The imagef in the bottom right of figure \ref{fig:gpd:supply:1} shows a histogram of supply events greater than zero and the corresponding linear fit. Table \ref{tab:gpd:supply} provides the maximum likelihood estimator (MLE) fitted parameters, and the standard error associated with each fitted parameter. Overall the fits are remarkably good: the combined data is clearly Pareto, as is the supply of grocers, who are the predominant type of donor in our data. The fits for manufacturers, individuals, and farms are not as strong, but still seem to comport to the POT model. This is an exciting result because it means that food supply (waste) events can be modeled with the same models used to predict and forecast extreme weather events. 

The values in table \ref{tab:gpd:supply} also reveal the distinct donation behavior of each category. Grocers
are fairly consistent donors, with a rate of 0.302 indicating that a typical grocery store donates some food on about 30\% of days,
with a mean weight of 369 lbs. Farms, on the other hand, donate infrequently with a rate of 0.023, which means that they donate
on about 2\% of days. However, when the donate, the mean quantity is much larger---around 6,000 lbs.

\begin{table}
\small
\begin{center}
\begin{tabular}{|l|c|c|c|c|c|c|}
\hline
Data & Threshold ($\theta$) & Rate ($r$) & Location ($\mu$) & Scale ($\sigma$) & Shape ($\zeta$) & Mean \\
\hline
All & 0 & 0.121 & 0 & 275.947 (5.382) & 0.439 (0.016) & 491.884 \\
Grocers  & 0 & 0.302 & 0 & 293.139 (6.130) & 0.205 (0.016) & 368.728 \\
Manufacturers & 0 & 0.038 & 0 & 562.549 (42.979) & 0.107 (0.051)& 629.954 \\ 
Individuals & 0 & 0.029 & 0 & 141.755 (18.374) & 0.905 (0.126) & 1492.042 \\
Farms & 0 & 0.023 & 0 & 918.811 (188.314) & 0.867 (0.200) & 6908.353\\
\hline
\end{tabular}
\caption{GPD fit parameters for daily supply and demand distributions in pounds. Standard Error values for the fitted parameters are given in parenthases.\label{tab:gpd:supply}}
\end{center}
\end{table}

\subsection{Predicting Supply}
\label{sec:predsupp}

Using the fit parameters in table \ref{tab:gpd:supply}, Pareto-distributed daily supply values can be generated:

\begin{equation}
s_{i,t} = \left\{\begin{array}{ll}
\mu + \frac{\sigma*(U(0,1)^{-\zeta} - 1)}{\zeta} & U(0,1) <= r\\
\theta & o.w.\\
\end{array}\right.
\end{equation}

\noindent Where $U(0,1)$ is a uniformly distributed random number in $[0,1]$. We use this function in our experiments to generate random, correctly distributed supply values that can be used in Monte Carlo\footnotemark style simulations. 

\footnotetext{Monte Carlo methods are ones that compute or predict the behavior of a (possibly) complex system by releated random sampling. In the classic approach, random samples are drawn from the appropriate probability distribution and then used in a deterministic algorithm to simulate the underlying dynamics of the system stochastically.}

In order to use this model to predict the supply of donors for which we do not have data, it is necessary to understand how other variables contribute to the shape and magnitude of the supply distribution. To this end, we investigate several other variables that might be correlated with mean daily supply: store size (building square footage), municipal zoning category, and store distance from the warehouse. We calculated zoning categories and square footageusing publicly available geographical information published by the City of Boulder \cite{eMaps}. The distance of each donor from the warehouse was calculated using drive-distance from the MapQuest Directions API \cite{mapquest}.

We perform a factorial analysis of variance (ANOVA) to see how each of these factors (individually and in combination) affects the mean daily supply. The GPD mean daily supply\footnotemark ($\bar{s}$) for a given supplier is:

\begin{equation}
\bar{s} = \mu + \frac{\sigma}{1 - \zeta}
\end{equation}

\footnotetext{This same analysis can be performed using well known parametric statistics (somewhat questionably), such as the arithmetic mean of daily supply. In that case the ANOVA and correlation results presented below still hold, but with slightly less confident p-values.}

\noindent To obtain this value for each supplier, we perform a GPD fit to determine the GPD parameters and then calculate the mean. The result of the ANOVA shows that the most important correlating variables are size and category. The relationship between donor size and
mean daily supply appears to follow a power law. An ANOVA gives F-values\footnotemark of 69.042 and 27.841 for $log_{10}$ of store square footage and donor category respectively, and 29.548 when used together. A Pearson's product-moment correlation test helps confirm this
relationship by finding a statistically significant correlation between the log of size and the log of mean daily supply for both grocers
alone (p-value $= 0.009$ and $\rho = 0.342$), and all suppliers together (p-value $= 0.097$ and $\rho = 0.413$). Given this, we can say that the mean daily supply (waste) and variability is independently correlated with the size of the donor. This is an important result because it allows us to predict the supply (waste) distribution for a given donor based simply on publicly available information: square footage and municipal zoning category.

\footnotetext{The F-value is a statistic that describes the ratio between explained variance and unexplained variance---or, put
differently, the ratio of between-group variability to within-group variability.}

\begin{figure}
\begin{centering}
\includegraphics[width=\columnwidth]{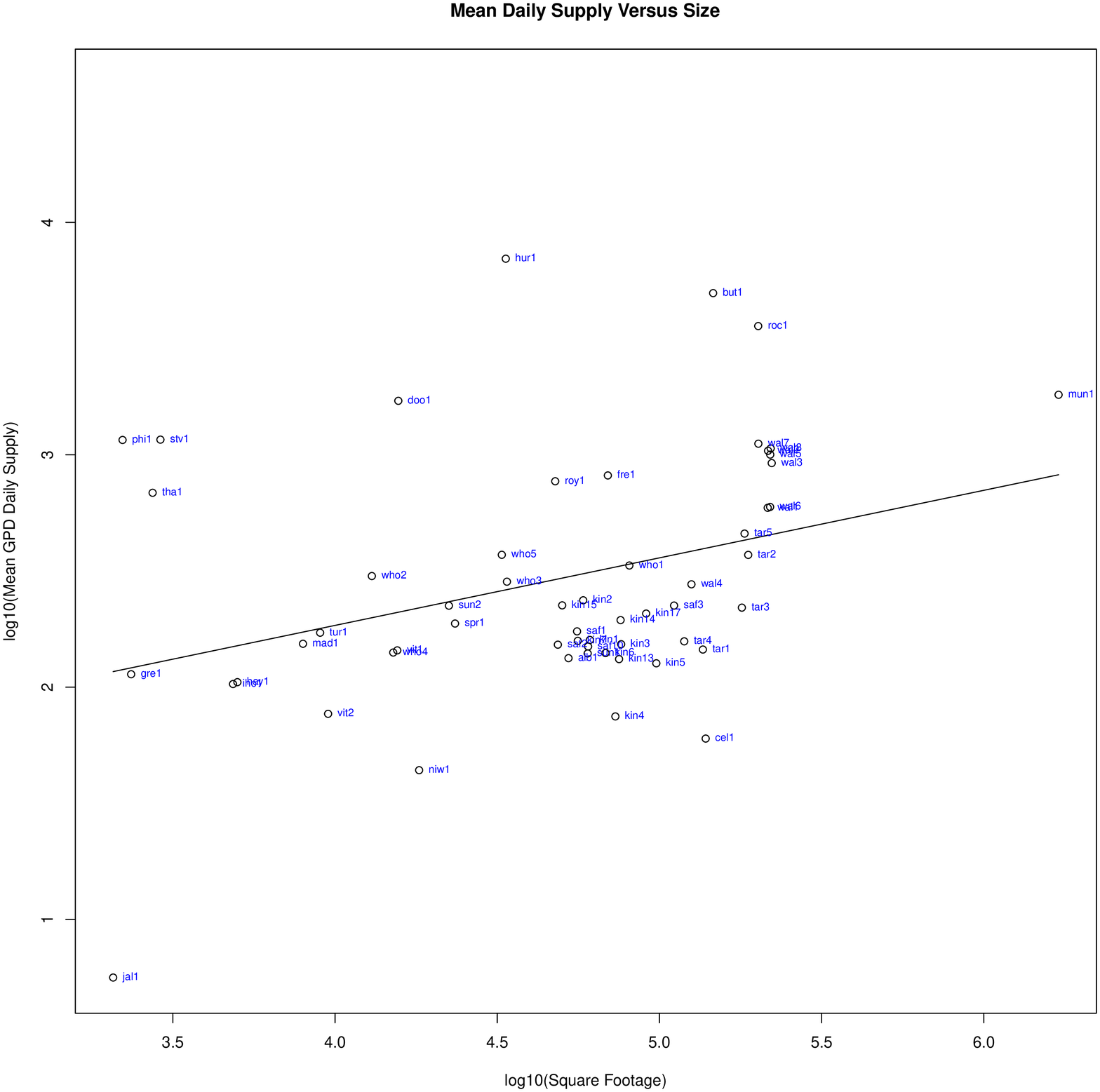}
\caption{Correlation between GPD mean and square footage\label{fig:sqftcorr}}
\end{centering}
\end{figure}

We perform a linear least squares fit to the log of GPD mean daily supply as a function of the log of donor square footage. This produces a fit with slope ($m$) of 0.291, intercept ($b$) of 1.103, and adjusted $R^2$ of 0.101. Figure \ref{fig:sqftcorr} shows this relationship along with the fit line. Clearly this fit is somewhat noisy, but after excluding outliers, such as phi1, stv1, doo1, hur1, but1, and roc1 that donate much more than other businesses of their size, and jal, niw1, and cel1, which donate much less than businesses of the same size, the relationship appears fairly solid.

Using this information we are able to predict the scale of the supply distribution for a new donor using its square footage ($x$):

\begin{equation}
log_{10}(\mu + \frac{\sigma}{1 - \zeta}) = log_{10}(mx + b)
\end{equation}

\noindent We know that the most important parameter is scale ($\sigma$), which by itself is well correlated with supply. We focus on predicting it and draw the remaining unknowns ($\mu, \zeta$) from the general fits given above. Solving for $\sigma$, and substituting in a constant $\zeta$ from table \ref{tab:gpd:supply}, 0 for $\mu$, and the constant fitted $m$ and $b$ values given above we obtain:

\begin{equation}
\sigma = 10^{log_{10}(x^m (1 - \zeta)) + b}
\end{equation}


\subsection{Characterizing Demand (Hunger)}
\label{sec:data:demand}

\begin{figure}
\begin{centering}
\subfigure[PDF]{\includegraphics[width=0.49\columnwidth]{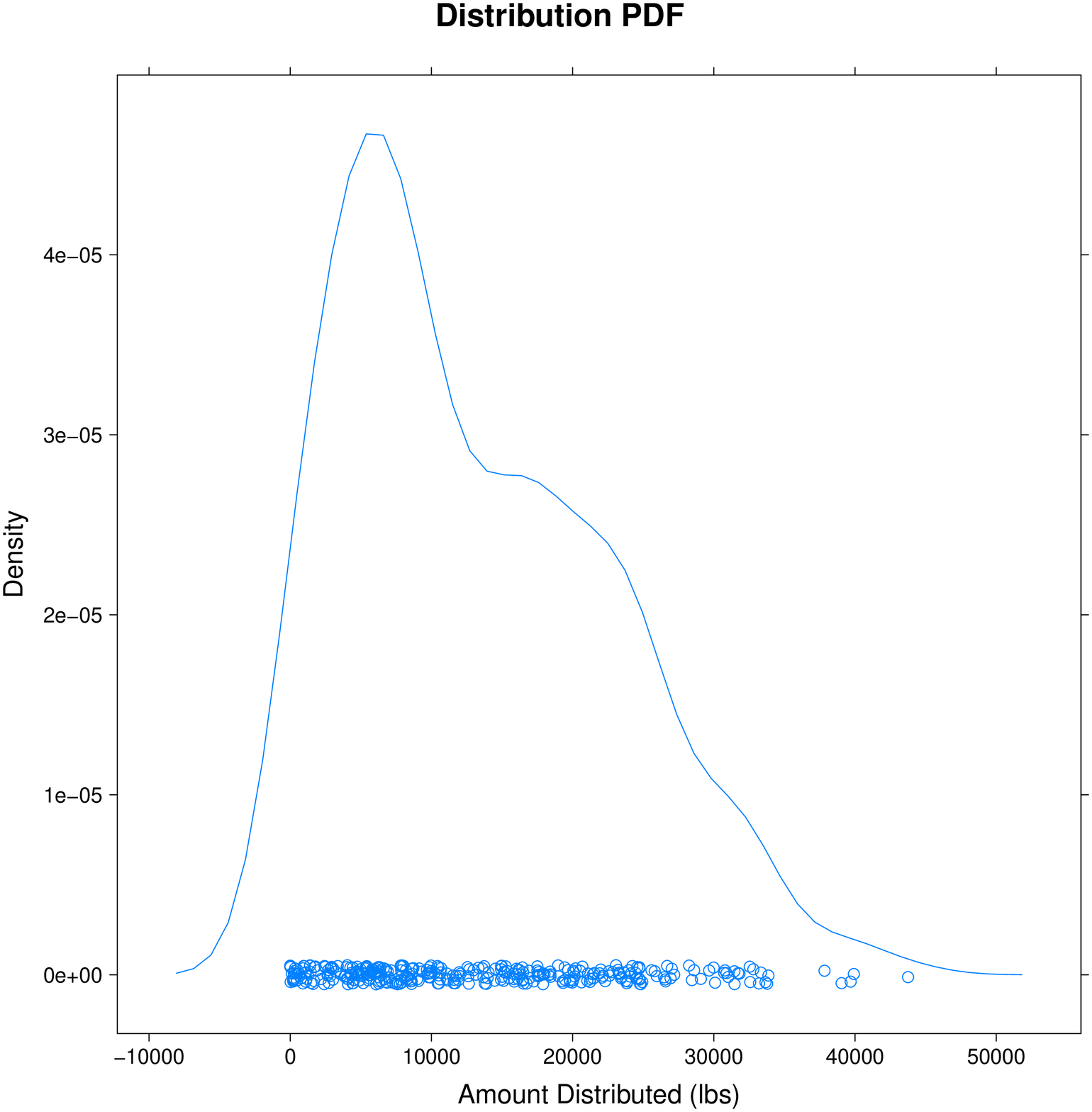}}
\subfigure[Normal QQ-Plot]{\includegraphics[width=0.49\columnwidth]{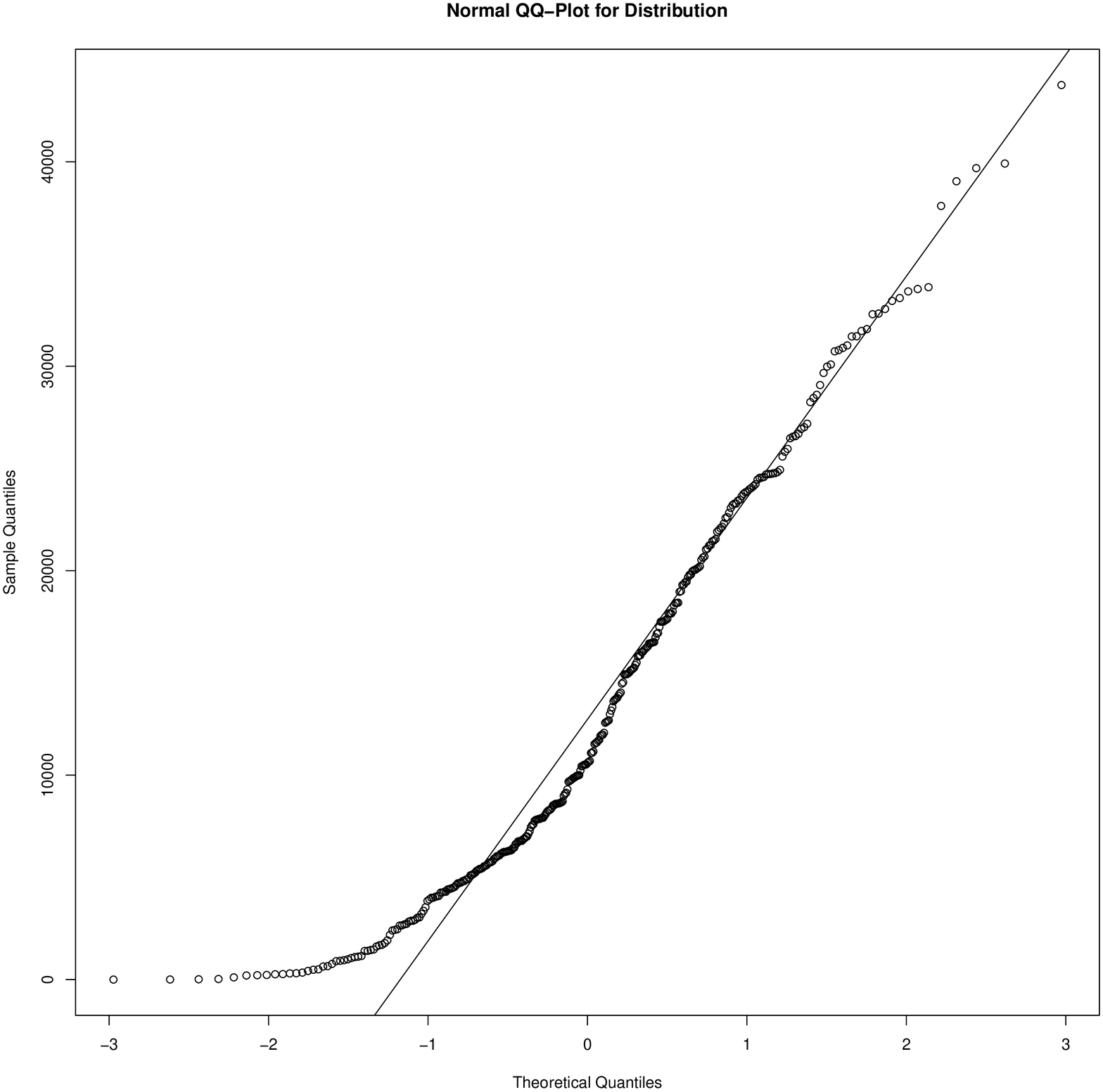}}
\caption{Daily total distribution data in pounds.\label{fig:dist:pdf}}
\end{centering}
\end{figure}

A final modeling task is to fit demand data. Due to privacy concerns for some agencies, it is not possible to determine the per-agency
demand from any defining characteristics, such as agency size or surrounding population density. Instead, we fit the aggregate daily demand, which corresponds to the amount of food delivered by CFS or picked up at the warehouse directly by the agencies. 

During the period for which we have data, CFS distributed food on 294 of 427 days, totalling 4,445,071 lbs. This total distribution amount is approximately twice that donated, since CFS purchases approximately 50\% of the food they distribute. Figure \ref{fig:dist:pdf} shows the kernel-smoothed probability density of a given daily demand (in lbs) for those 294 days. The distribution on any given day when food is distributed appears to be Gaussian, as is shown by the Q-Q plot given in figure \ref{fig:dist:pdf}. The assumption that the underlying distribution is Gaussian seems reasonable given that total demand is likely a strict function of the number of hungry people, which should not be highly varying in time. Hence, we can summarize the distribution with the mean (10,410 lbs) and standard deviation (10,041 lbs). We take this as the demand for food in pounds per day that CFS currently serves.

This empirically derived average value is not a perfect indicator of the actual demand in the CFS service area. The food required to feed all people with food insecurity could actually be much higher. Several reports have detailed food insecurity in the US, reaching different conclusions about the extent of the problem. According to \cite{FeedingAmericaReport}, which describes the efforts of the Feeding America program, 5.7 million unique individuals (or 1.661\% of the US population\footnote{The US Census Bureau estimates the population of the US as 307,006,550 in 2009.}) are served each week by the approximately 37,000 agencies aligned with their program. There are 348,017 individuals in the service area of CFS\footnote{The US Census Bureau estimates the population of Boulder and Broomfield counties at 303,482 and 55,990 respectively.}, meaning that, based on national-level statistics, there are approximately 5,781 unique individuals in this region per week who need food assistance. In \cite{Nord2009}, Nord et al. show that nationally in 2009, 14.7\% of households were food insecure at some point during the year, 9\% had low food security, and 5.7\% had very low food security\footnote{Low food security is defined by the USDA as ``Reports of reduced quality, variety, or desirability of diet. Little or no indication of reduced food intake'' and very low food security is defined as ``Reports of multiple indications of disrupted eating patterns and reduced food intake'' \cite{USDALabels}.}. Using the 5.7\% figure, that would indicate that 20,490 individuals have very low food security in the service region of CFS. A local study conducted by Feeding America in conjunction with CFS found that approximately 10,800 unique  individuals seek food assistance per week in the CFS service region \cite{CFSFeedingAmericaReport}. Using USDA statistics for average consumption of food, a typical American in 2010 consumes approximately 2.85 lbs of food per day, of which the majority is meat and protein (0.41 lbs) and grain (0.48 lbs) \cite{USDAFoodStats}\footnote{This assumes that the weight of dairy products is equivalent to the same volume weight of water and the weight of vegetables and fruit is equivalent to half the weight of the same volume of water. The latest available usage statistics are from 2008.}. Given this, if we assume that each individual who has very low food security acquires a third of their daily intake via CFS, the mean daily demand would be between 5,491 lbs (using national Feeding America levels\footnote{This value is remarkably close to the mean 5,454 lbs currently donated to CFS each day.}), 10,260 lbs (using local CFS Feeding America service statistics\footnote{This number is remarkably close to the mean 10,410 lbs already distributed by CFS daily.}), 19,465 lbs (using USDA very low food security percentile), and 48,600 lbs (using USDA low food security perentile). This is a staggeringly large range that serves to highlight the fact that consensus on hunger and food demand does not exist. 



In the remainder of this paper, we treat the high end estimate of 48,600 lbs per day as the gold standard---the equivalent of providing one meal to every person in the service area who experiences some form of food insecurity. Although this would equate to feeding people who, for the most part, have enough money to feed themselves, it would do a great deal towards helping the poorest Americans avoid weighing tradeoffs between nutritional value of purchased foods, and other expenses, while at the same time substantially reducing waste sent to landfills. Similarly, we will treat 10,260 lbs as the low end goal, which equates to the net demand that CFS and other area agencies already attempt to fulfill during times of highest demand (i.e., a week when all unique clients come to pick up food). In an average week, CFS is already distributing this much food (10,410 lbs), but typically only rescues half of it, and purchases the remainder. It will be the primary aim of our investigation to determine the feasibility, and quantify the cost, of rescuing all the food necessary to meet this demand goal.


In the next section we describe the remainder of the model, which defines the mechanics of supply and demand using an optimization framework. In section \ref{sec:implementation}, we tie together that optimization model and the statistical models in this section into a simulator that can be used to study the dynamics and feasibility of the food redistribution problem.





\section{Model}
\label{sec:model}

In this section we develop a model of food supply, demand, and cost of delivery between areas of supply and areas of demand. The premise of the model is simple: there are some number of potential donors that generate some amount of recoverable waste each day. Performing a pickup at a given donor incurs a cost, which is related to how far away that donor is, and optionally how much food is to be transported. The goal of the model is to determine a schedule for each day that chooses which donors to visit such that food demand (hunger) is met and minimizes the cost of doing a pickup. 

We begin by discussing optimization problems in general and related approaches in order to help motivate the direction we have chosen. Then, we describe the general model for our problem and two important extensions to improve its realism that take into account food expiration and food storage at a central warehouse.

\subsection{Background and Related Problems}
\label{sec:background}

Optimization is the process of finding the best option from a set of options using a fitness function to capture the relevant details of the problem \cite{Papadimitriou1982}. In the food redistribution problem, the relevant details are the food supply from donors, food demand from agencies, and the cost of picking up food from donors and delivering it to agencies. 

There are several classes of problems that provide context for the food redistribution problem, and these classes are not mutually exclusive. There are resource allocation problems in which limited resources are allocated to maximize a goal, such as profit, i.e. \cite{Holloran1986,Roy1982}. There are also minimum cost flow problems such as shortest-path, transportation, or maximum flow network, i.e. \cite{Tang2006,Sharkey2008}. In shortest-path problems, the objective is to find the minimum distance, time, or cost for a sequence of activities or travel through a network. The maximum flow problem expands on shortest-path by adding constraints on the flow capacity between nodes, which limits the flow to some maximum value. In real-world examples, this would be equivalent to restricting the number of cars on a road between cities, or the size of pipelines. In transportation problems, the objective is to determine the minimum cost between centers of supply and demand. 

With food distribution, the objective is to find the minimum distance through a network of donors that provides enough food to meet demand. In this way, it is a shortest-path problem, where the flow has to be greater than a constraint, which is the demand. A fitness function can be formulated that is used in a solution method. One common method for solving these types of problems is linear programming (LP) \cite{Hillier2005}. This approach finds the optimal solution on problems where inputs and constraints are linear combinations. Mileage, for example, is linear. In food redistribution, the cost per mile is the same regardless of which path in the network is being travelled. In addition, the decision to visit any given store on that day is discrete --- we either go to the store or we do not. If we visit the store, we get 100\% of the supply available, but we also incur 100\% of the cost of travel. 

While LP is common for solving minimum flow problems, it is possible for these problems to also be nonlinear. In food redistribution for example, we can use the LP approach to find the optimal solution using a daily cost calculation. However, formulating the problem to find a multi-day pickup schedule introduces non-linearities that make the LP approach less effective. The multi-day solution generates a pickup schedule based on knowledge of food supply of the current and subsequent days as well as how the food expires over time. The most cost-effective solution could be to only pickup food once a week for any given store. However, if the store has perishable items, one week might be too long to wait. In addition, if a store has a small donation one day, but will have a large donation the next day, the optimal solution might be to batch the pickups---skipping the pickup on the current day and picking up both the current and the next day's supply at one time. 


For complex optimization problems that cannot be solved using LP or other simple solving methods, heuristic and metaheuristics methods are often used to find approximate solutions \cite{Hillier2005}. These approaches have been used for minimum flow, i.e. \cite{Oldham2001,Fleischer2002} and resource allocation problems in which there is a time element that requires a multi-day solution, i.e. \cite{Christou1999,Gorman1998}. Determining whether metaheuristic methods can be used to find efficient multi-day schedules is an area we expect to investigate in future work.

\subsection{Food Redistribution Model}

Our model is as follows. We start with $N$ donors with food supply, and $M$ agencies with food demand\footnote{The units used here are not important, and could be pounds, calories, or any other reasonable metric to quantify food.}. The aggregate available food supply ($\hat{s}$) on a given day ($t$) is the sum of the supply from each individual donor:

\begin{equation}
\hat{s}_t = \sum_{i}^{N} s_{i,t} \\
\end{equation}

\noindent Similarly, aggregate demand ($\hat{d}$) on a given day ($t$) is the sum of demand at each individual agency:

\begin{equation}
\hat{d}_t = \sum_{i}^{M} d_{i,t} \\
\end{equation}

\noindent Picking up food everyday from every donor would retrieve every pound of available food, but it might not be cost effective. Instead, we establish a schedule where everyday, there is a pickup from a subset of donors. This multi-day pickup schedule is controlled by a boolean matrix ($r_{i,t}$), which identifies which suppliers have pickups scheduled on which days: 

\begin{equation}
r_{i,t} = \left\{\begin{array}{ll}
1 & pickup\;at\;supplier\;i\;on\;day\;t\\
0 & o.w.
\end{array}\right. \\
\end{equation}

\noindent Each donor is associated with a constant cost of making a pickup ($c_i$). The total cost ($\hat{c}_t$) on day ($t$) is the cost to visit the selected donors on that day: 

\begin{equation}
\hat{c}_t = \sum_{i}^{N} c_i*r_{i,t}\\
\end{equation}

\noindent The total supply for that pickup schedule ($\hat{q}_t$) is then:

\begin{equation}
\hat{q}_t = \sum_{i}^{N} s_{i,t}*r_{i,t}\\
\end{equation}

\noindent In this formulation, the objective is to minimize $\hat{c}_t$, such 
that $\hat{q}_t \ge \hat{d}_t$ for $\forall t$. This will minimize cost and provide enough supply to meet demand:

\begin{equation}
min \; \sum_{t}^{T} \hat{c}_t\;s.t.\;\hat{q}_t \ge \hat{d}_t
\end{equation}

\subsection{Food Expiration}

Establishing a multi-day pickup schedule for food involves a unique challenge in that food can go bad. Some of the food not picked up on day $t$ will expire by day $t+1$, but the other food will remain. The rate of expiration is related to the way the food is stored, the state of the food, and the type of food. In this model we have made the simplifying assumption that all food expires at the same rate regardless of these conditions. The food supply available on day $t+1$ is the new supply for that day, plus the previous day's supply that was not picked up and did not expire:

\begin{equation}
s_{i,t+1} = s_{i,t+1} + \epsilon(1)*s_{i,t}*(!r_{i,t})
\label{eq:supplyrec}
\end{equation}

\noindent where $\epsilon(\Delta t) \in [0,1]$ quantifies the fraction of food not expected to expire over $\Delta t$ nights. $!r_{i,t}$ is the inverse of the boolean scheduling matrix (and hence is 1 when a pickup does not occur and 0 otherwise). Further extensions can consider more complete definitions of $\epsilon$. The recurrence in equation \ref{eq:supplyrec} can be trivially rewritten as a sumation of the previous $t$ days:

\begin{equation}
s_{i,t}' = s_{i,t} + \sum_{a=0}^{t-1} \epsilon(1)*s_{i,a}*(!r_{i,a})
\end{equation}

\subsection{Food Storage}

Another important component in the model is a central warehouse where excess food can be stored
after it is picked up. This allows for overages in recovery to be used the following day. In our model, the warehouse supply 
on a given day $t$ is the amount picked up on day $t$ minus the day $t$ demand plus the warehouse supply from the previous day that did not expire:

\begin{equation}
\hat{w}_t = (\hat{s}_t - \hat{d}_t) + \hat{w}_{t-1}*\epsilon(1)
\end{equation}

\noindent It should be noted that the assumption of a central warehouse is not universal. For instance, the largest
food rescue organization in the United States, City Harvest in New York City, operates without a central 
warehouse and instead delivers goods in the same day they are rescued, often as part of the same
tour \cite{cityharvest}. CFS, on the other hand, operates using a central warehouse in Niwot, Colorado. Each day, three full-time drivers make scheduled pickups at area donors and bring the goods back to the warehouse, where they are stored for later distribution.


\section{Implementation}
\label{sec:implementation}

In this section we tie together the empirical results in section \ref{sec:data} and the model outlined in section
\ref{sec:model} to develop a simulation framework that can be implemented and studied in practice. In particular, we
discuss approximations and assumptions that are necessary in order to allow experimental analysis of the problem in a reasonable
amount of time.

\subsection{Iterative Multiday Solution}

Finding the optimal solution using the model described in section \ref{sec:model} is combinatorial and has a worst-case complexity
of $O(2^{NT})$ for $N$ suppliers and $T$ days. For all but the smallest $N$ and $T$, this problem is intractable. However, as discussed in section \ref{sec:background}, the optimal solution for a single day, with constant costs, can be formulated as a linear program and solved quickly:

\begin{eqnarray}
\hat{c} = c_1*r_1 + c_2*r_2 + \cdots + c_n*r_n\\
min \; \hat{c}\;s.t.\;\sum_{i}^{N} r_i*s_i \ge \sum_{i}^{M} d_i
\end{eqnarray}

\noindent A sub-optimal solution for multiple days can be calculated by applying this linear program iteratively, once for each day:

\begin{equation}
C = \sum_{t}^{T} \; min \; \hat{c}\;s.t.\;\sum_{i}^{N} r_i*s_i \ge \sum_{i}^{M} d_i
\end{equation}


The main
failing of this approach is that it is unable to make multi-day decisions, where a day's pickup schedule is based on knowing the supply for the following days. However, in
the real world, it is unlikely this information would be available anyway. In this sense, the iterative solution may be more
realistic than the the general purpose optimal solution.

For each time step (day), an optimal solution is found using a linear program (LP) solver. The LP searches for a schedule ($r$) that minimizes the total cost subject to the constraint that the amount recovered must be greater than or equal to the amount demanded. To solve this equation on a computer we look to the freely distributed \textit{lp solve} project, which is able to solve mixed integer
linear programs \cite{lpsolve}. Because our solution is actually a binary vector, the mixed integer and binary extensions are necessary. An example problem, presented in the lp language used by \textit{lp solve} is given in figure \ref{fig:lpex}.

\lstdefinelanguage{lp}{morekeywords={min,c1,bin},sensitive=true}
\lstset{frame=trbl,captionpos=b,float}
\begin{figure*}
\begin{lstlisting}{label=fig:lpex,language=lp}
min: 24.39x0 + 56.59x1 + 35.81x2 + 46.01x3 + 24.64x4 + 29.62x5 + 36.654x6;
c1: 177.99x0 + 552.06x1 + 12.0x2 + 5.2x3 + 2.1x4 + 3.02x5 + 1.2x6;
bin x0 x1 x2 x3 x4 x5 x6;
\end{lstlisting}
\caption{Example \textit{lp solve} program formulated in the lp language.\label{fig:lpex}}
\end{figure*}

\subsection{Determining Costs}

Constant costs for each supplier must be calculated in order for this problem to be solved as a linear program. The data from CFS provides us with the names of 90 suppliers. We first manually determine the street address of each donor and then calculate the distance of each supplier from the CFS warehouse. 

One of our research
goals is to understand the supply available at candidate donors that are not currently working with CFS. Therefore, we have attempted
to locate all grocers, bakeries, and manufacturers in Boulder and Broomfield counties that might supply donations. The total list of
donors is 156. We cannot claim that this list is exhaustive, but we believe it represents the bulk of grocers, bakeries, and food manufacturers in the region. In future work we hope to include restuarants and cafes as well, but have excluded them in the present
study because CFS does not pick up at these locations.

To calculate distance, we use the MapQuest Driving API directions \cite{mapquest}. For each supplier in the CFS data, we retrieve driving directions (which presumably use an optimized shortest path, taking into account the actual constraints of the roads) and use the total driving length of the first offered route, in kilometers, as the cost of visiting that supplier.

\subsection{Clustering Suppliers}


\begin{figure}
\begin{centering}
\includegraphics[width=\columnwidth]{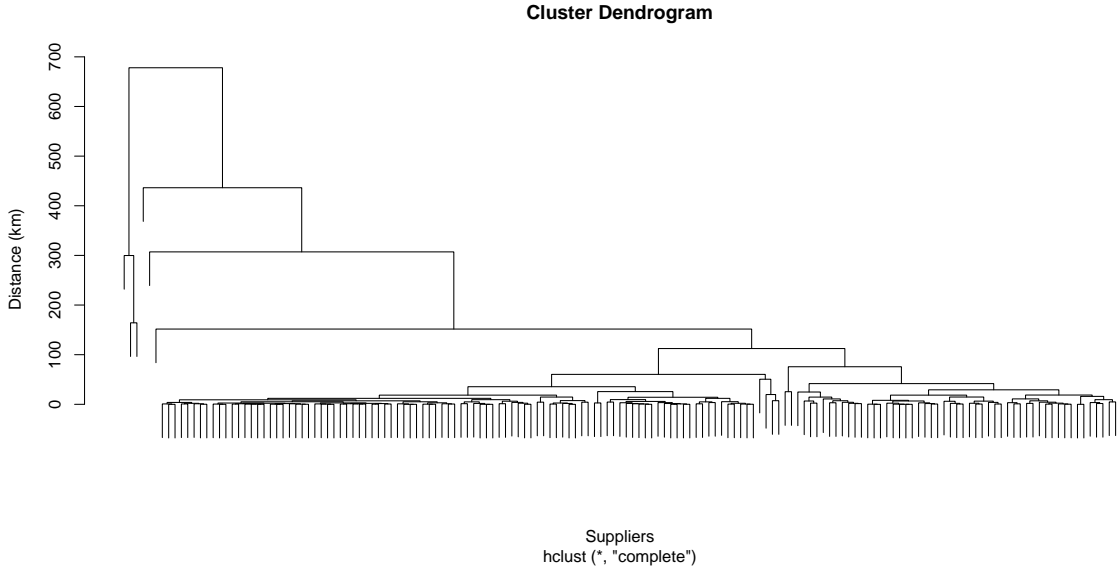}
\caption{Hierarchical clustering of suppliers based on distance.\label{fig:distclust}}
\end{centering}
\end{figure}

In order to allow for batching pickups at nearby locations, we determine how our 156 suppliers are clustered together. We use the MapQuest Driving API to calculate shortest-route driving distance between all pairs of nodes. This requires 12,012 unique queries and takes some time\footnotemark. Figure \ref{fig:distclust} shows a higherarchical
dendrogram of all 156 suppliers based on distance to each other. We proceed with a k-means clustering
with a goal of $k = 50$ clusters when considering all 156 suppliers and $k = 30$ clusters when considering the smaller set of 90. Although somewhat arbitrary, this keeps the size of clusters roughly constant (3 members per cluster on average) in both data sets. 

\footnotetext{Originally, we made use of the Google Maps API, but eventually abandoned it in favor of MapQuest because of rate limit restrictions and daily request quotas.}

In k-means clustering, observations are partitioned into $k$ clusters so that each observation belongs to the cluster whose center (mean location) is nearest. To solve this, we use the method of Hartigan and Wong, as implemented in the R statistics package \cite{Hartigan1979,R}. The result is 50 points in space that the 156 suppliers are centered around so that the mean distance from any supplier to their cluster point is minimized. 

When using the resulting computed clusters, we make the basic assumption that when a pickup is done at any cluster member, all the other cluster members should also be visited. In this sense, each supplier is its own cluster and the supply at that cluster is the sum of all members.
To determine the cost of visiting a cluster we take the average driving distance of all cluster members from the central warehouse, multiply it by two (to count the return trip), and then add it to the minimum sum of the distances from one node to each other node. Hence, the cost for visiting a given cluster $k$ is:

\begin{equation}
c_k = \frac{2}{N_k} \sum_i^{N_k} c_{w,i} + min_j \sum_i^{N_k} c_{i,j} \; s.t. \; i,j \in \mathbf{cluster}_k
\end{equation}

\noindent where $N_k$ is the number of members in cluster $k$, $c_{i,j}$ is the cost of driving from $i$ to $j$, and $w$ is the index of the warehouse.

Although we cannot claim that this is a perfect solution, it is quick to compute, reduces the computational complexity of the problem, and models the way food rescuers actually do pickups. As we will see,
it also produces cost values that are very close to the average costs incurred by CFS. Hence, we proceed with this cost function
for the remainder of the paper.

\subsection{Implementation Outline}

Algorithm \ref{alg:sim} shows the pseudocode implementation of the complete simulation algorithm. For each day, the total demand ($\hat{d}$) is a constant based on the expected demand as discussed in Section \ref{sec:data:demand}. If we are considering food storage in a warehouse and there is food already available, this total demand is reduced by that amount to achieve the remaining net demand. Without considering storage, the net and total demand will always be equal. 

Supply values for each participating donor are drawn stochastically from the GPD distribution derived in Section \ref{sec:data:supply}.
The scale parameter is calcuated for each donor using the square footage, as discussed in section \ref{sec:predsupp}. Determining the square
footage for each location is a laborious task due to inconsistent and somtimes unavailable geographic information system (GIS) data.
For those suppliers for which it is possible, we compute the square footage from municipal data for Boulder County \cite{eMaps}. For those suppliers where this data is unavailable, we use GIS tools and orthoimagery to calculate square footage manually.

Supply accumulates each day at donors if it is not recovered. However, it also expires at a constant rate of $1 - \epsilon$. In our implementation, $\epsilon$ is a constant between 0 and 1, but in future work we hope to investigate more realistic epsilon calculation (perhaps by type of food). 

In some cases, there is simply not enough supply to meet demand, in which case there is no feasible solution to the LP. When this occurs, we assume that all suppliers must be visited to get as much as possible (whatever the cost).

\begin{algorithm}
 \caption{Multiday Food Redistribution Simulator}
  \label{alg:sim}
  \begin{algorithmic}[1]
    \State $D \gets$ constant demand
    \State $\epsilon \gets$ constant epsilon value
    \State $seed \gets$ seed for random number generator
    \State $\mathbf{rate} \gets$ constant rate from GPD fit for each category
    \State $\mathbf{shape} \gets$ constant shape from GPD fit for each category
    \State $\mathbf{scale} \gets$ constant shape from GPD fit for each category
    \State $\mathbf{scale'} \gets$ predicted or categorical scale for each supplier
    \State $\mathbf{cats} \gets$ category of each supplier

    \State $loc \gets$ constant location from GPD fit
    \State $m \gets$ constant slope of linear log/log relationship between mean supply and square footage
    \State $b \gets$ constant intercept of libear log/log relationship between mean supply and square footage
    \State $w \gets 0$

    \State $\mathbf{sqft} \gets$ square footage of each supplier
    \State $\mathbf{c} \gets$ constant supplier costs
    \State $\mathbf{olds} \gets 0$

    \For{$i = 1 \to N$}
      \If{$sqft_i$ is available}
        \State $scale'_i \gets 10^{log_{10}(sqft_i^m (1 - shape_{cats_i})) + b}$
      \Else
        \State $scale'_i \gets scale_{cats_i}$
      \EndIf
    \EndFor
 
    \For{$t = 1 \to T$}
      \For{$i = 1 \to N$}
        \State $r_1 \gets$ random value in $[0,1]$
        \State $r_2 \gets$ random value in $[0,1]$
        \If{$r_1 \le rate_{cats_i}$}
          \State $s_i \gets loc + scale'_i * \frac{r_2^{-shape_{cats_i}} - 1}{shape_{cats_i}}$
        \Else
          \State $s_i \gets 0$
        \EndIf
        \State $s_i \gets s_i + olds_i * \epsilon$
      \EndFor

      \State $w \gets w * \epsilon$
      \State $d \gets D - w$

      \State $C,\mathbf{r} \gets lpsolve(\mathbf{c},\mathbf{s},d)$     
      \State $S \gets \sum_{i}^{N} r_i * s_i$

      \If{$d - S > 0$}
        \State $w \gets w + (d - S)$
      \EndIf
      \For{$i = 1 \to N$}
        \State $olds_i \gets (!r_i) * s_i$
      \EndFor
    \EndFor
  \end{algorithmic}
\end{algorithm}


\section{Experiments}
\label{sec:experiments}

One contribution of this paper is evaluating the feasibility of the redistribution approach for alleviating hunger. In this section we describe a study using the model from section \ref{sec:model} and stochastically generated data from the empirical distributions described in section \ref{sec:data} to create simulations of food redistribution scenarios. 

Our objectives with these experiments are to:
\begin{itemize}
\item Demonstrate that the model works in that it reproduces observed food supply from the CFS data.
\item Evaluate how model parameters, such as $\epsilon$ and warehousing, affect food availability.
\item Determine the relationships between food availability, the number of donors, and cost.
\end{itemize}

The approach we take here is a classic repeated measures approach, where the average behavior of a complex system is studied through repeated Monte Carlo simulation and \textit{ex post facto} analysis. Each simulation is run for one year (365 days) and uses the same random seed (so that results can be compared). By studying the problem via simulation, we are able to investigate both the average case and extreme scenarios of food redistribution, which serves to answer whether this approach fills the gap between demand and current supply. 

In this first set of experiments, we only consider pickups from the 90 donors that are already participating with CFS. The goal here is to reproduce and understand the dynamics of the food redistribution problem: how much energy (cost) must be expended to meet the worst-case demand, whether demand can be met reliably, how frequently underruns ($supply < demand$) and overages (excess pickup) occur, and how the rate of food expiry limits the amount of recoverable food. We include the warehouse, since CFS operates with a warehouse. We use an $\epsilon$ of 0.5, indicating that 50\% of ``waste'' food is expected to expire in 24 hours (or, put another way, half of the food will remain on day $t+1$, one quarter on day $t+2$ and so on). Then, using the same set of suppliers, we evaluate how food availability changes as a function of cost and $\epsilon$. These experiments show that removing cost restrictions affects how much food could be retrieved, as well as how food expiration contributes to food availability.

Next, we extrapolate the supply that could be available from businesses not currently donating. We include all 156 donors in the CFS distribution area. We examine how cost changes for a fixed number of donors by simulating a range of demand goals. We also examine the problem from the other angle --- how supply changes for a fixed cost as the number of donors changes. Presumably, as more potential donors participate, the travel distance between donors decreases, which lowers the cost for pickups. In all cases, we also evaluate the role of $\epsilon$ in food supply by simulating an $\epsilon$ of 0 to 1. This includes the full range of food expiration time scales in our model. In these studies, we use a demand of 10,260 lbs per day, which is the demand from the CFS Feeding America service statistics and represents the lower bound on the food required to fully meet the worst case demand in the CFS service area, as described in section \ref{sec:data:demand}.  

\section{Results}
\label{sec:results}

\begin{figure}
\begin{centering}
\subfigure[Demand and Recovery]{\includegraphics[width=0.9\columnwidth]{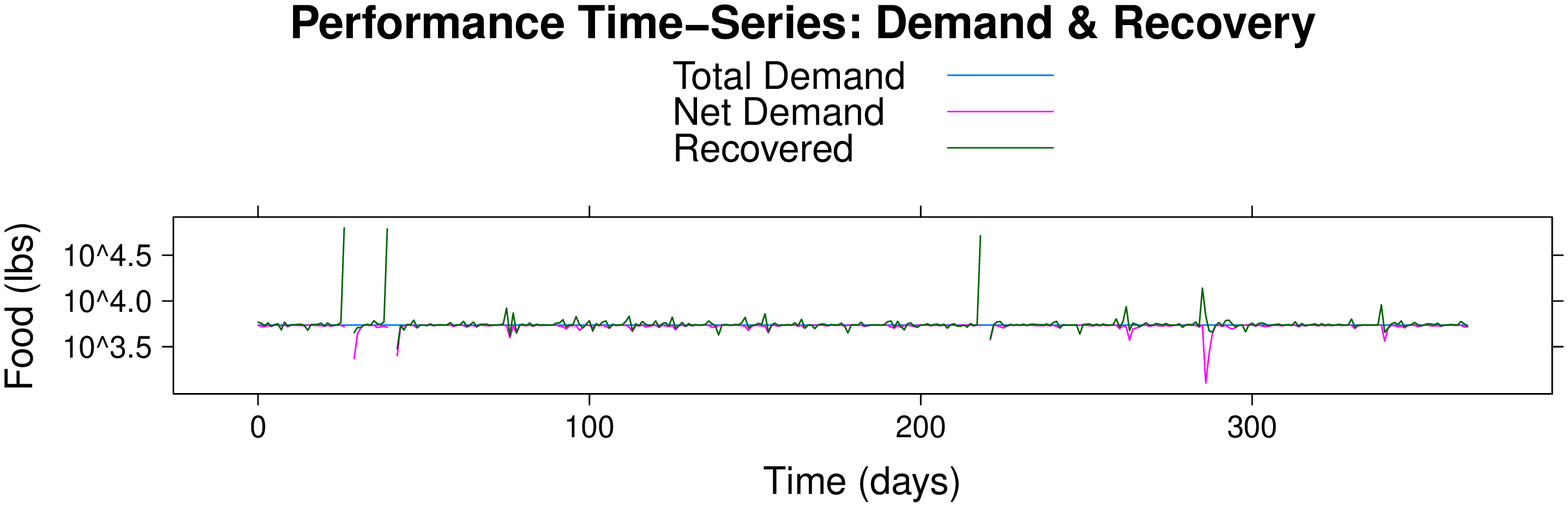}}

\subfigure[Cost]{\includegraphics[width=0.9\columnwidth]{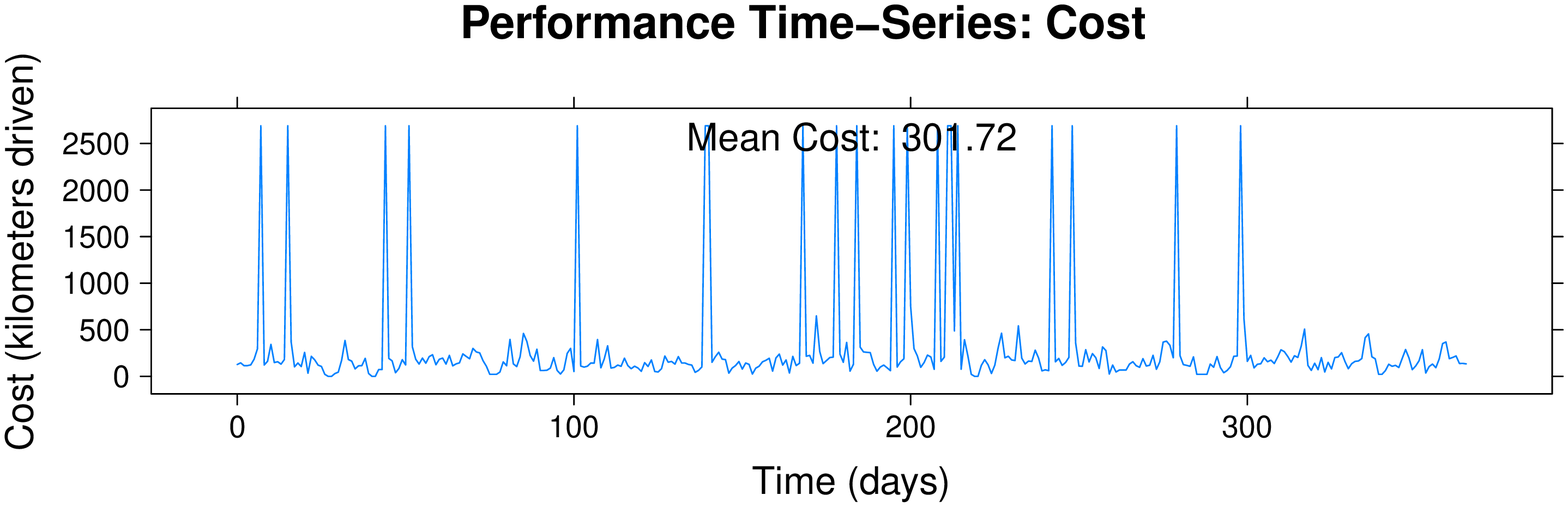}}

\subfigure[Excess]{\includegraphics[width=0.9\columnwidth]{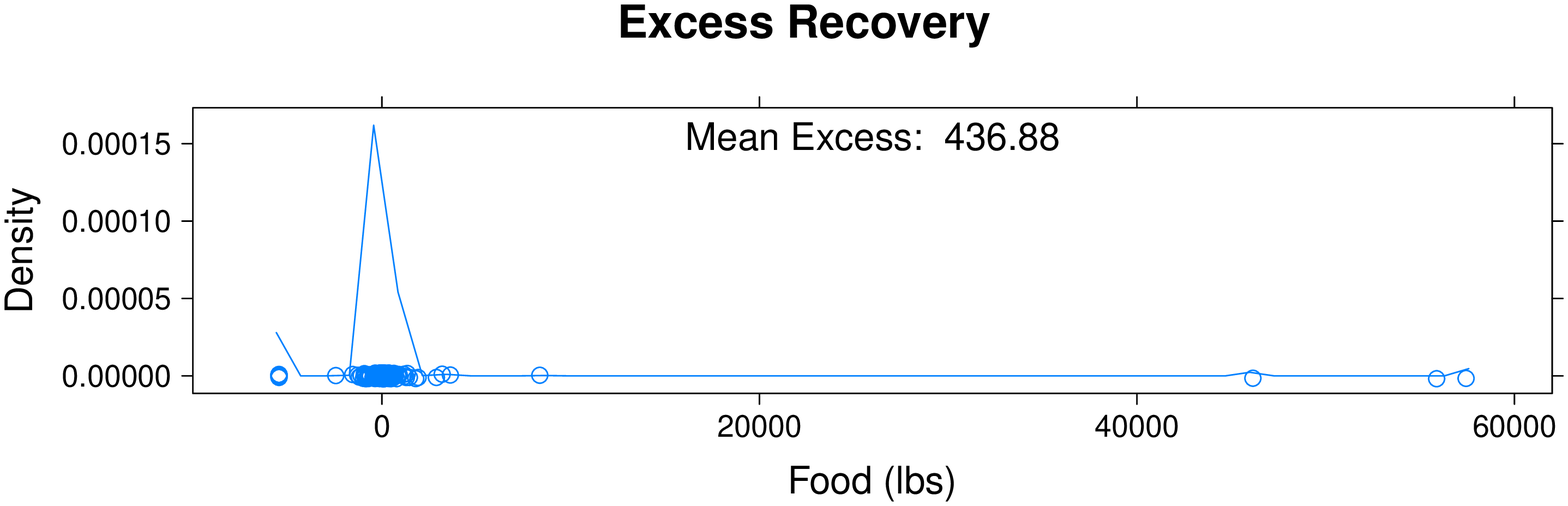}}
\caption{Performance time-series for one year simulation using the existing CFS donors, $\epsilon = 0.5$, a target demand of 5454 lbs, and using a central warehouse. Excess is the difference between net demand and recovered. Hence, a positive excess indicates demand has been met.\label{fig:results:basic}}
\end{centering}
\end{figure}

Figure \ref{fig:results:basic} provides a first look at the simulation-based results using parameters intended to reproduce the 
CFS model. Here, we set the daily demand to 5,454 lbs, the mean received by CFS in our data set. The top plot shows the supply and demand.
The total demand is fixed at 5,454, the net demand is each day's demand reduced by the amount present in the warehouse, and the ``recovered'' line is the amount recovered for each of 365 days. On the majority of days, the demand is met. The bottom plot shows the distribution of excess, where a positive excess reflects an overrage and a negative excess reflects a shortage in food recovered. The excess appears to be normally distributed about the mean of 436.88 lbs, indicating that the recovered food from donors was generally higher than the demand. This amount is driven up by spikes of food rescue, which occur at three times during the year. These correspond to extremely large random food rescue events, which are also observed from farms and manufacturers in the real data. Finally, the middle plot shows the cost on each day, which has a mean of 301.72 (kilometers driven). CFS estimates their daily driving distance (sum of three vehicles, without optimizing paths) to be 212 km \cite{tomreed}. The difference here stems from the fact that our model counts the cost of picking up food at distant farms and manufacturers, which generally deliver food directly to CFS themselves \cite{tomreed}. If we exclude donors farther than 100 km away, the mean cost drops
to 198.35 km, which is within 10\% of the value provided by CFS, without producing any more underruns than would be seen otherwise.	

\begin{figure}
\begin{centering}
\subfigure[Demand and Recovery]{\includegraphics[width=0.9\columnwidth]{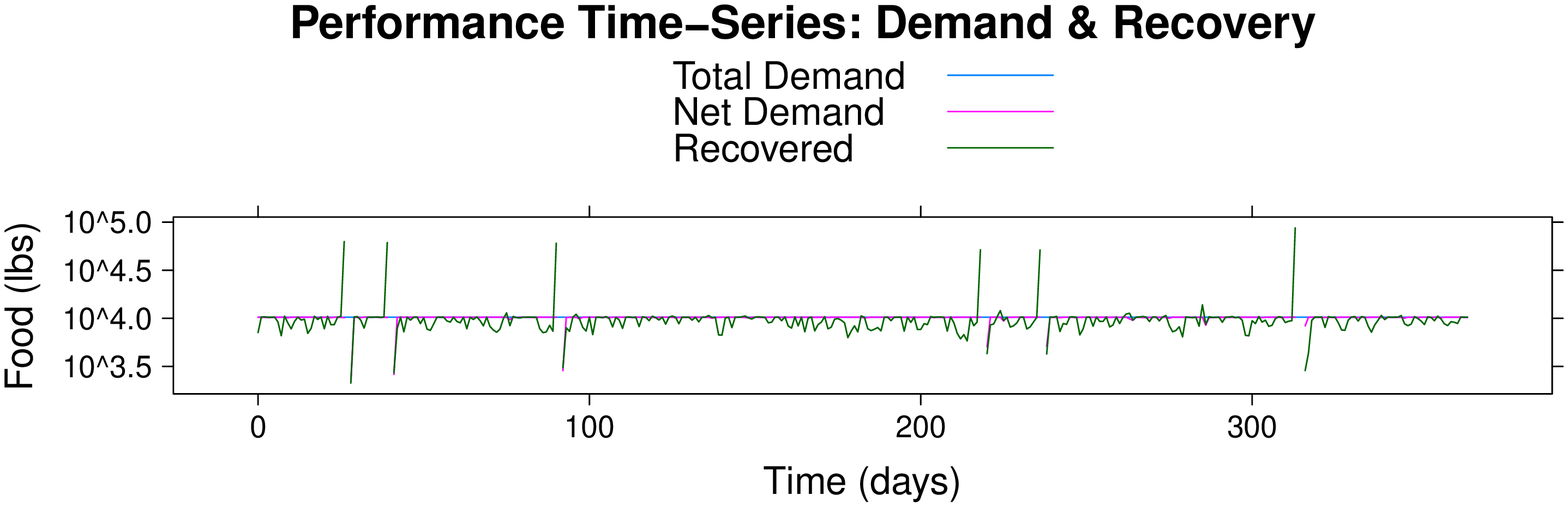}}

\subfigure[Cost]{\includegraphics[width=0.9\columnwidth]{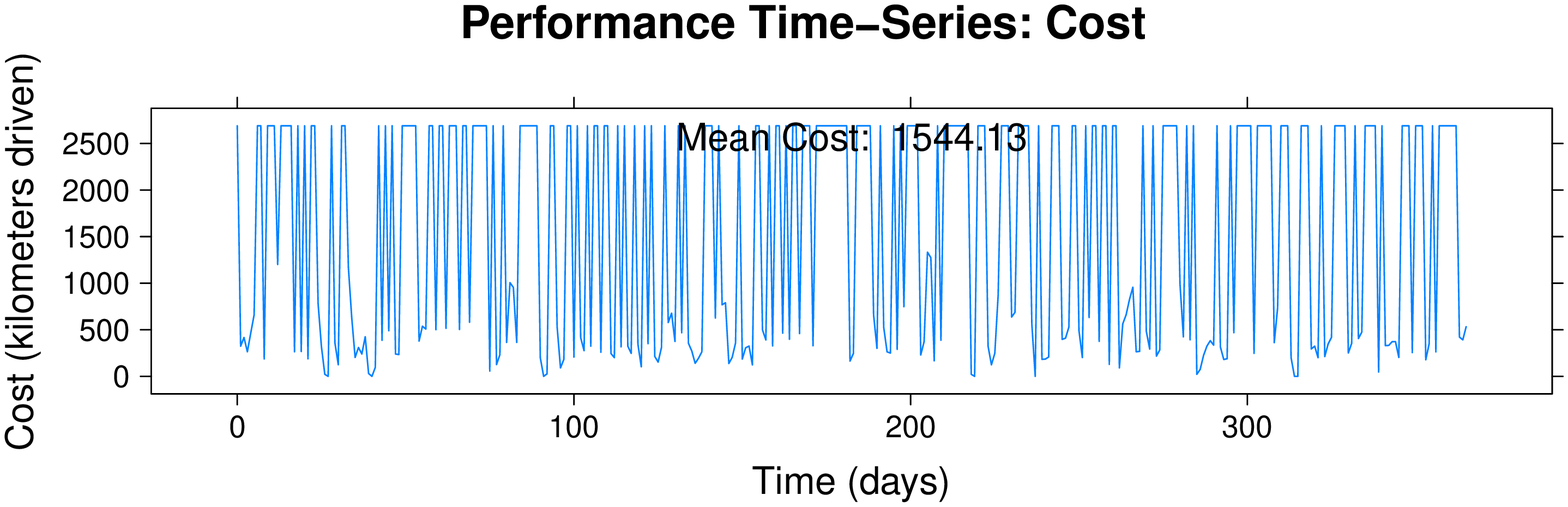}}

\subfigure[Excess]{\includegraphics[width=0.9\columnwidth]{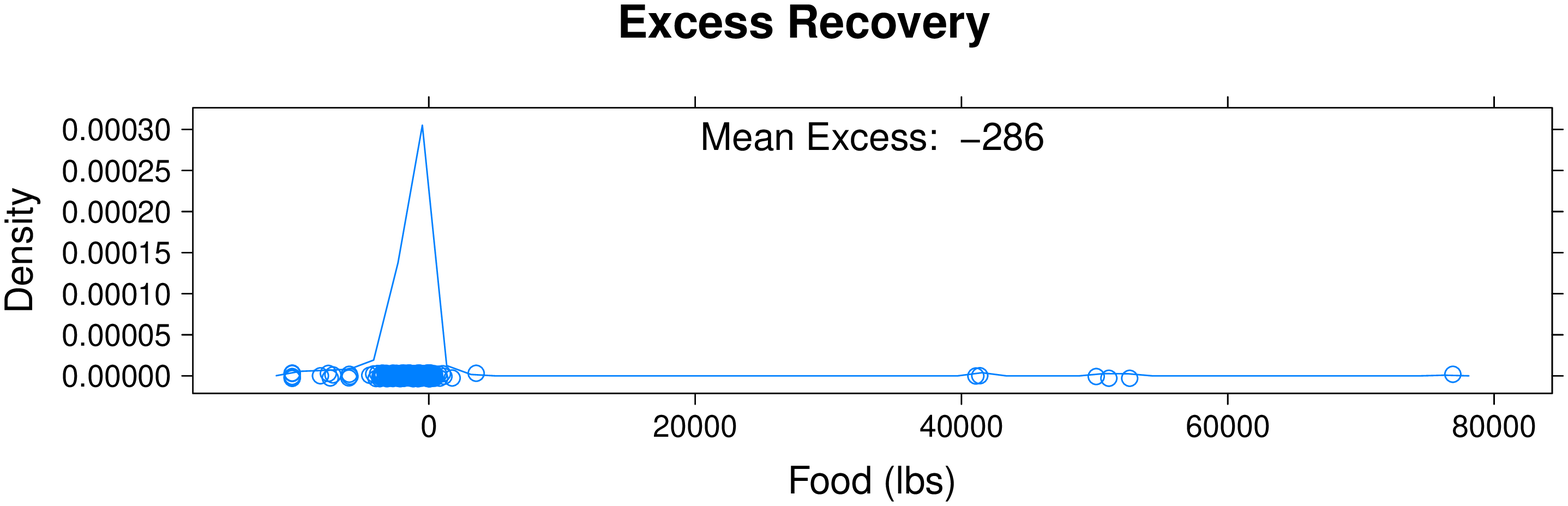}}
\caption{Performance time-series for one year simulation using the existing CFS donors, $\epsilon = 0.5$, and using a central warehouse. The target demand is set at the larger rescue threshold of 10,260. Excess is the difference between net demand and recovered. Hence, negative excess values indicate a shortage.\label{fig:results:basic10k}}
\end{centering}
\end{figure}

Using the goal of 10,260 lbs/day and $\epsilon=0.5$, underruns are the norm rather than the exception, as shown in Figure \ref{fig:results:basic10k}. A mean cost of 1,544.13 km is necessary in order to meet this demand, which is an average 286 lbs shy of the demand each day. If we exclude donors greater than 100 km away, the mean cost drops to 703.1 km, but the average daily shortage (opposite of excess) decreases to 262.16 lbs. The observation that \textit{in some cases excluding the furthest away donors can substantially reduce cost} (by 54\% in this case) while obtaining approximately the same amount of food suggests that there may be some fundamental density of donors required for efficient food rescue---in scenarios where donors are sparsely distributed relative to recipients, the cost of rescue may be very high for the same amount of food. On the other hand, in denser environments our model can capitalize on the random supply from nearby (and clustered) donors to drive down cost. Understanding the effect of spatial distribution of donors (and density) is an interesting question for future work.

\begin{figure}
\begin{centering}
\subfigure[Demand and Recovery]{\includegraphics[width=0.9\columnwidth]{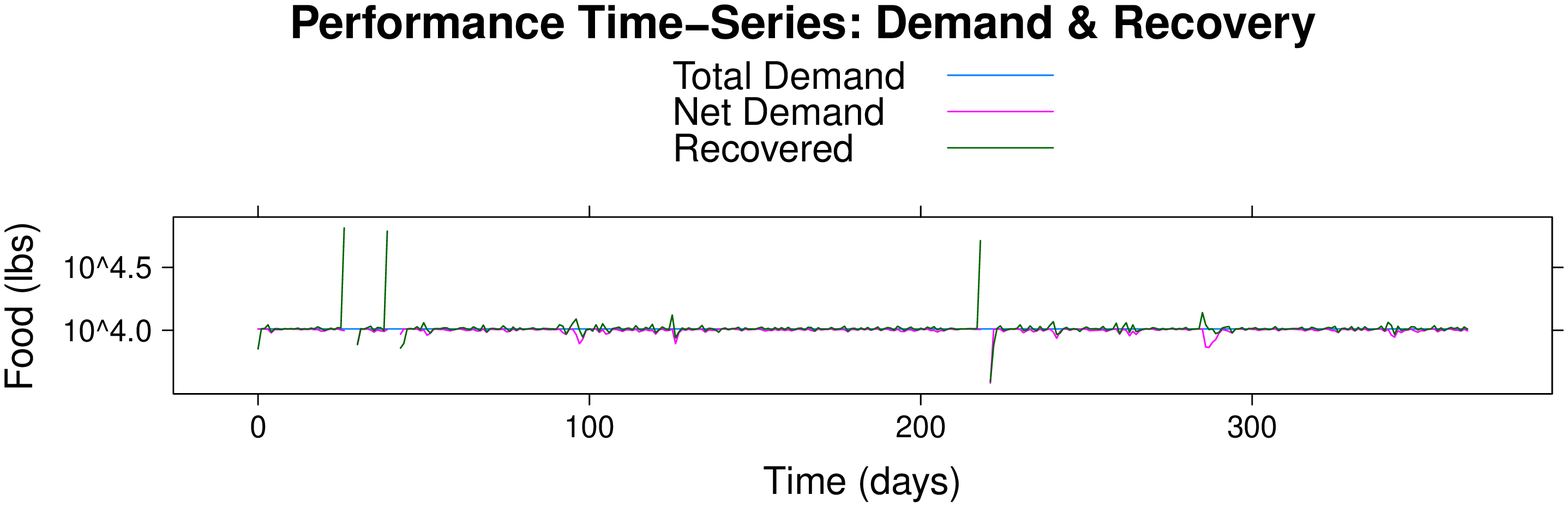}}

\subfigure[Cost]{\includegraphics[width=0.9\columnwidth]{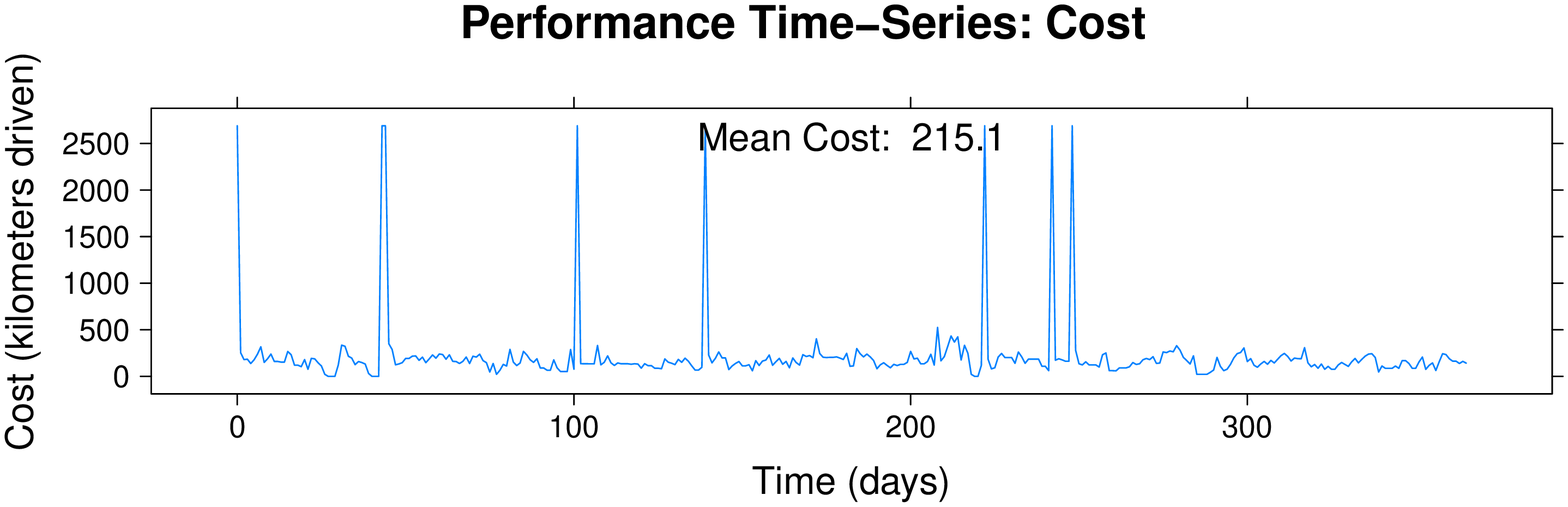}}

\subfigure[Excess]{\includegraphics[width=0.9\columnwidth]{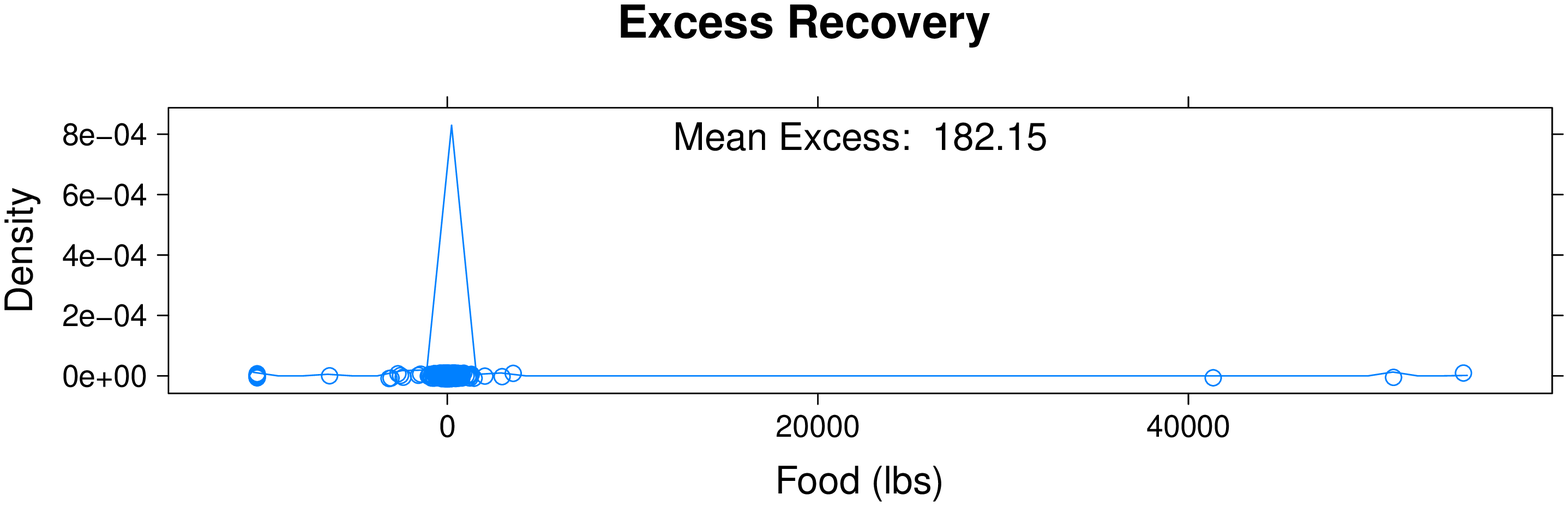}}
\caption{Performance time-series for one year simulation using the existing CFS donors, $\epsilon = 0.8$, and using a central warehouse. The target demand is set at the larger rescue threshold of 10,260. Excess is the difference between net demand and recovered. Hence, negative excess values indicate a shortage.\label{fig:results:basic10ke08}}
\end{centering}
\end{figure}

\begin{figure}
\begin{centering}
\includegraphics[width=0.7\columnwidth]{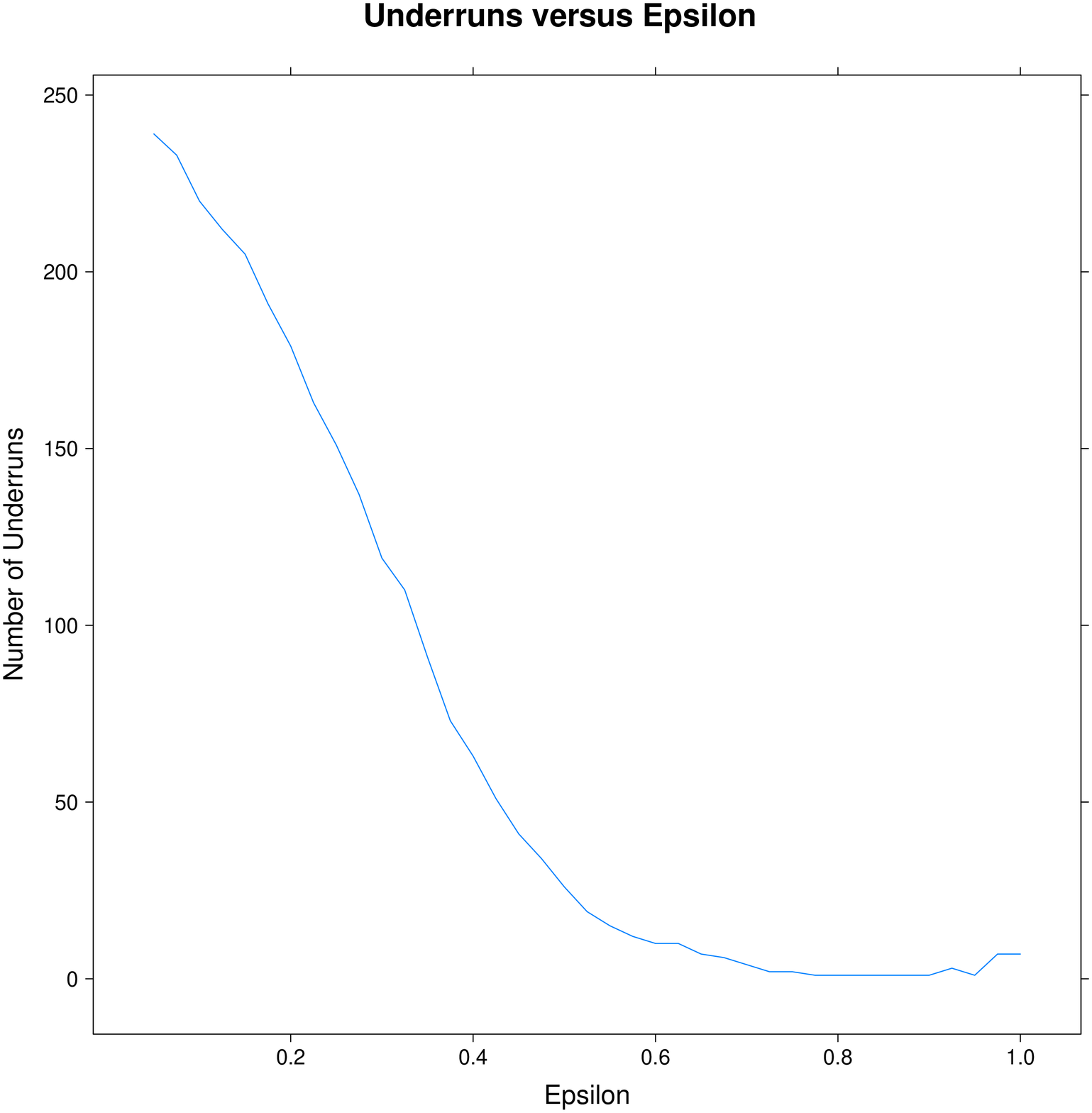}
\caption{Number of underruns (for a 365 day simulation) as a function of the epsilon value. Results are from using the set of 90 suppliers with a central warehouse. \label{fig:undereps}}
\end{centering}
\end{figure}

Another important feature of the model is the $\epsilon$ parameter. Figure \ref{fig:results:basic10ke08} shows the same scenario, but with $\epsilon = 0.8$, indicating that food expires at a rate of 20\%, instead of 50\% as above. In this case, demand is met on most days and the mean excess is 182.15, with an average cost of 215.1 km. This is nearly an eight-fold reduction in cost for the same amount of food, simply by changing the rate at which food expires. Figure \ref{fig:undereps} shows this relationship explicitly by plotting the number of days in a 365 day simulation where supply failed to meet demand as a function of the $\epsilon$ value. For small $\epsilon$ values, the number of underruns is very high; as $\epsilon$ is increased, meeting the higher demand becomes obtainable and the shortages are mitigated. Clearly, the effect of $\epsilon$ cannot be underestimated. In a practical sense, this highlights \textit{the importance of proper food storage in the recovery process to smooth supply and mitigate shortages}. Those involved in retail food distribution are well aware of the importance of keeping food in the optimal conditions from farm to shelf, but these results show that it is also an essential consideration in the viability of a food recovery model. For this reason, many food banks focus on non-perishable food items during large fundraising events and food drives. This can be understood in the terms of our model: canned goods have an epsilon near 1.0.  

\begin{figure}
\begin{centering}
\subfigure[Demand and Recovery]{\includegraphics[width=0.9\columnwidth]{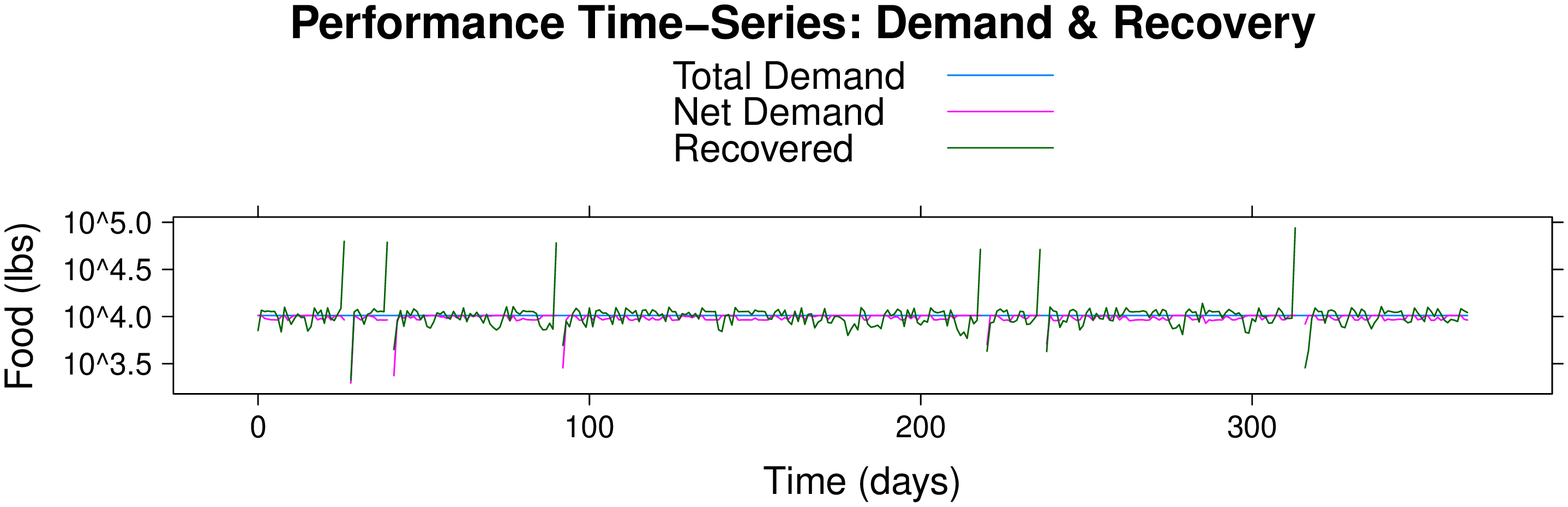}}

\subfigure[Cost]{\includegraphics[width=0.9\columnwidth]{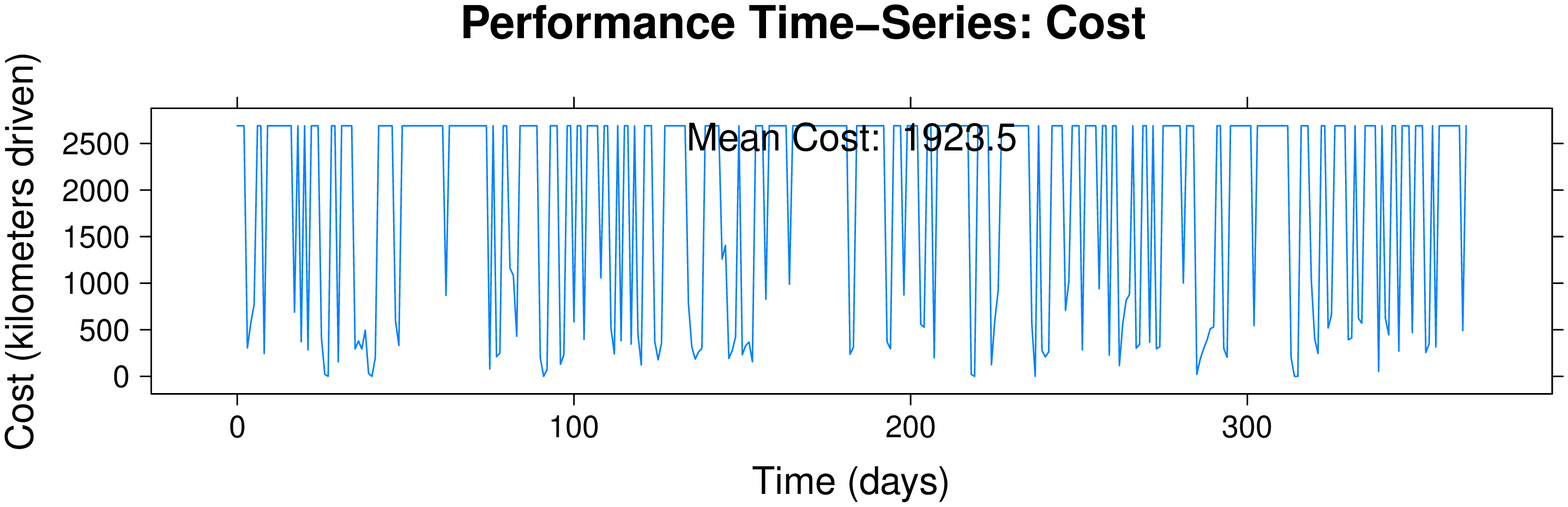}}

\subfigure[Excess]{\includegraphics[width=0.9\columnwidth]{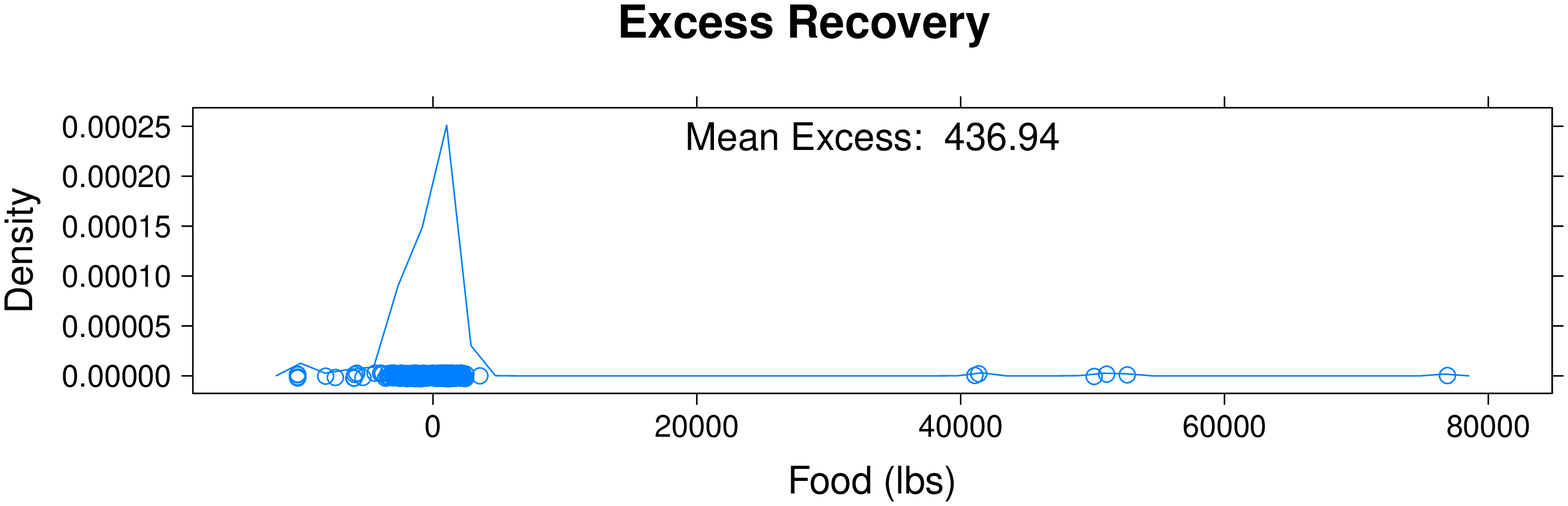}}
\caption{Performance time-series for one year simulation using the existing CFS donors, $\epsilon = 0.5$, and using a central warehouse. The target demand is set at the larger rescue threshold of 10,260. Excess is the difference between net demand and recovered. Hence, negative excess values indicate a shortage. In this version, we intentionally collect 20\% more food each day than we predict is necessary to attempt to smooth out underruns on subsequent days.\label{fig:results:basic10ko20}}
\end{centering}
\end{figure}

A final consideration of the basic model is intentional overage. Figure \ref{fig:results:basic10ko20}, shows the same scenario as above, with $\epsilon = 0.5$, but purposely attempting to retrieve 20\% more food each day than is actually required. The logic here is that we can capitalize on freak supply ``floods'' to prepare for potential ``droughts'' in supply. We can see that in this case, there is a marginal improvement in mean excess, and actually a small increase in cost. From this, we are able to say that this strategy is not an effective one---\textit{picking up extra food now cannot protect from shortages in the future}. Presumably this stems from the fact that food expires at the same rate whether it is left at the supplier, or transfered to the warehouse. In the case where that is not true (for instance, the supplier does not have room to store expiring ``waste'' for pickup and discards anything not picked up immediately), then this strategy may be more effective.

\begin{figure*}
\begin{centering}
\subfigure[90 Suppliers]{\includegraphics[width=0.6\columnwidth]{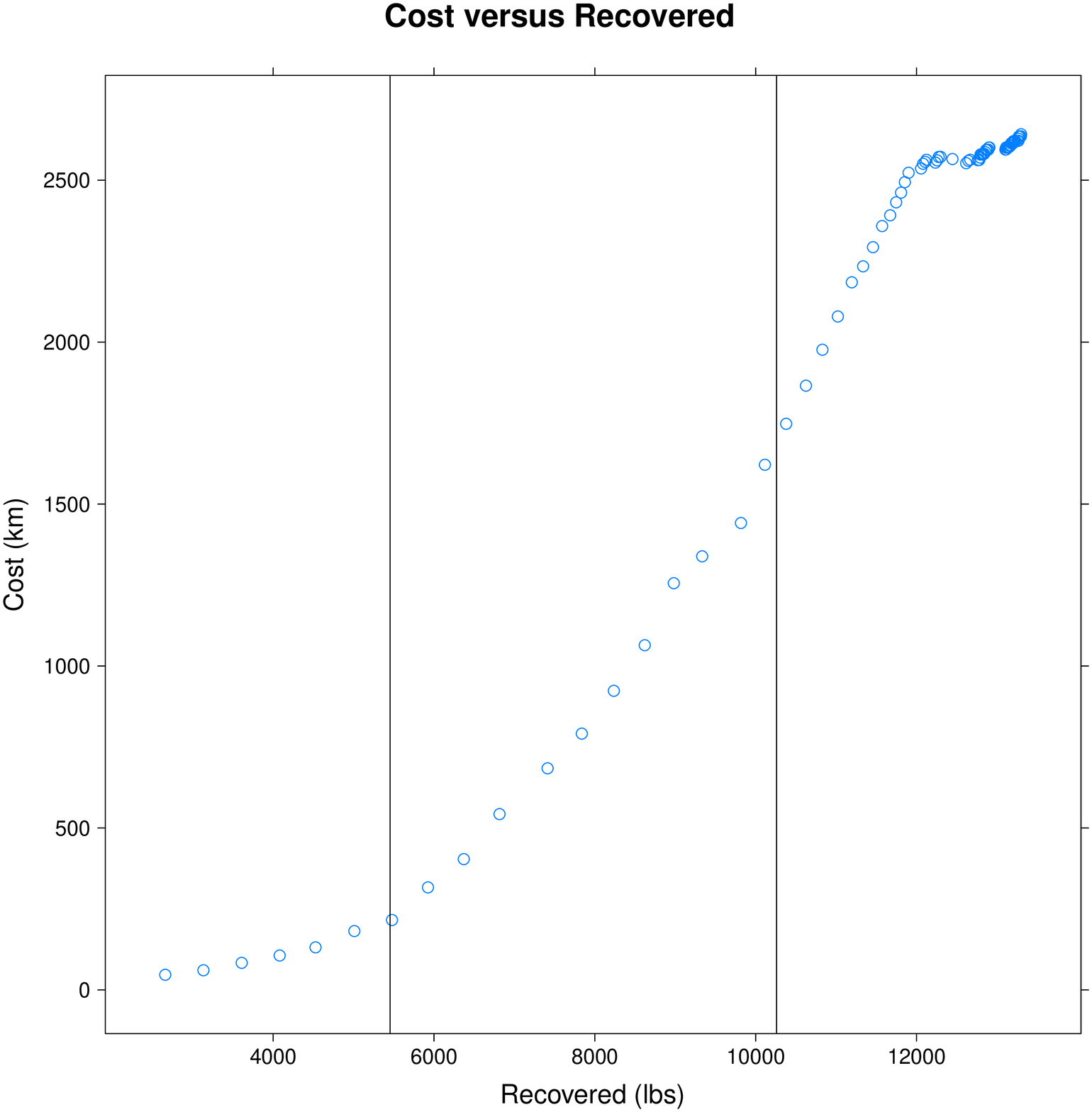}}
\subfigure[156 Suppliers]{\includegraphics[width=0.6\columnwidth]{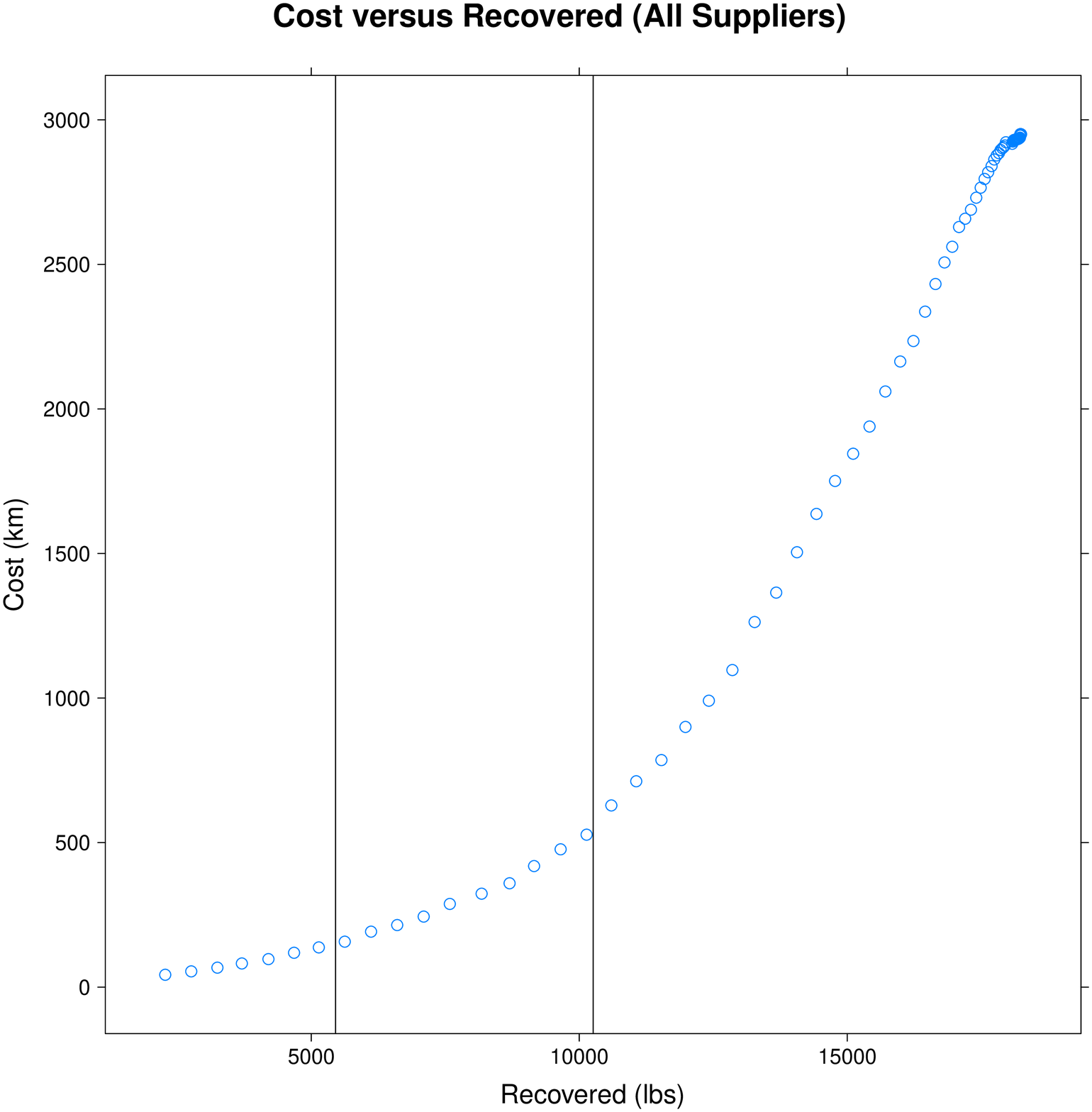}}
\caption{Cost of recovering food for each of two sets of suppliers. The vertical lines indicate demand reference points: 5,454 lbs and 10,260 lbs.\label{fig:costvrec}}
\end{centering}
\end{figure*}

Next, we consider the rest of the suppliers that are not currently donating food, but could be, and increase the total set of suppliers
from 90 to 156. Figure \ref{fig:costvrec} shows the cost vs. food recovery relationship for both the set of 90 suppliers
and the set of 156 suppliers. In both plots, there is a saturation point, where there is no more recoverable food. For the set of 90 suppliers
this maximum is around 13,000 lbs. For the larger set of suppliers, the maximum is closer to 19,000 lbs. There are two conclusions that can be drawn from this result. First, \textit{despite the complexity of the underlying model, the relationship between the amount of food available for rescue and the number of donors participating appears to be linear}. We quantify the benefit of increasing donor participation below. Second, and perhaps more importantly, with sufficient resources, and more participating donors, CFS may be able to comfortably meet their current worst-case demand without purchasing food. To meet this goal, they would only need to drive approximately 500 km a day, which is slightly more than two times their current expenditure. This indicates that \textit{provided sufficient funding is available, and a large number of businesses are participating as donors, the food rescue model can successfully feed the area's hungry using only food that would otherwise be wasted}. Admittedly, the demand of 10,260 lbs is well below the gold standard of 48,600. For that to succeed, according to our model, CFS would need to have sufficient resources to drive at least 3,000 km per day. 

\begin{figure}
\begin{centering}
\includegraphics[width=0.7\columnwidth]{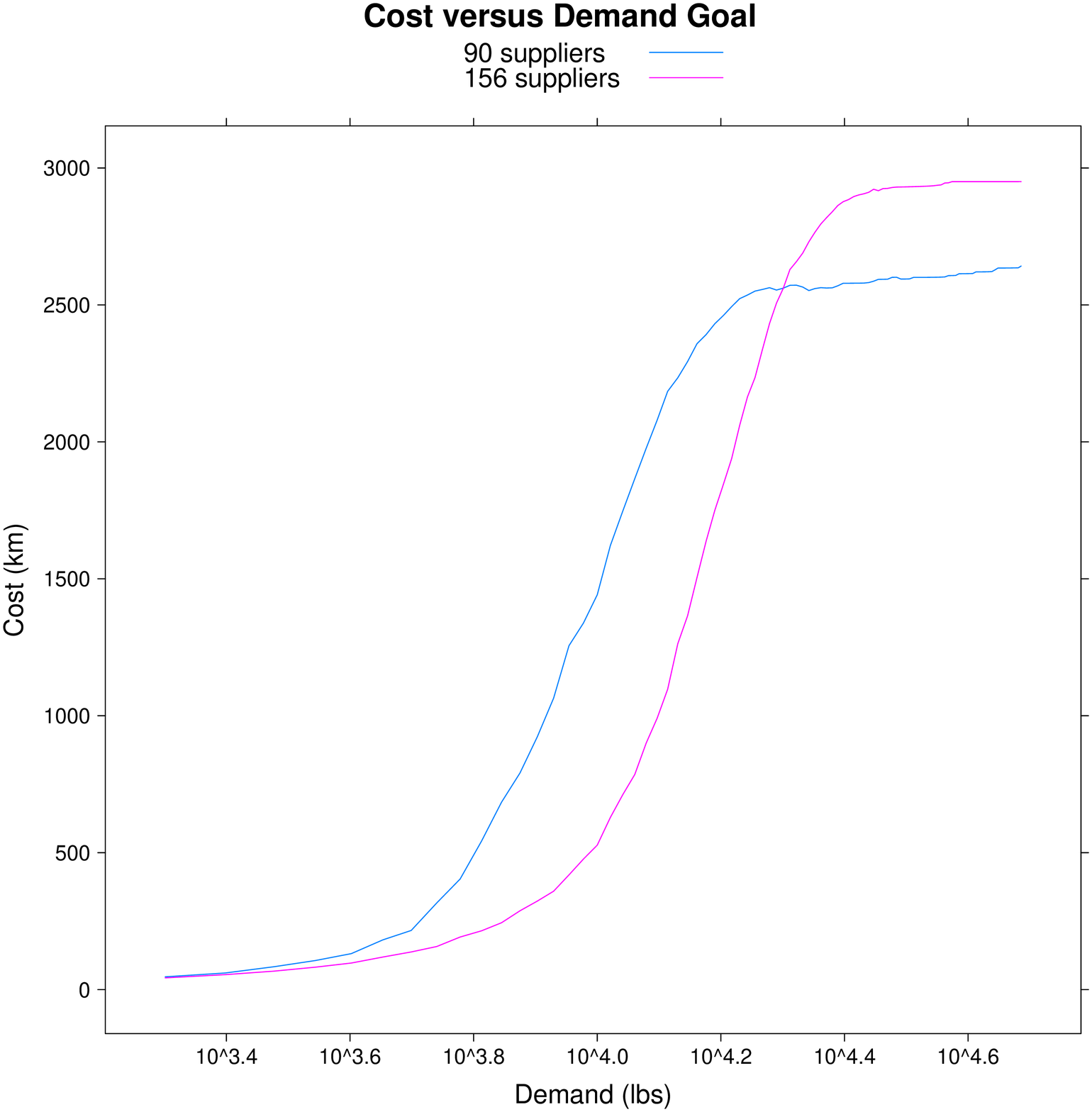}
\caption{Relationship between cost and demand goal for each of the supplier sets.\label{fig:costvdem}}
\end{centering}
\end{figure}

\begin{figure}
\begin{centering}
\includegraphics[width=\columnwidth]{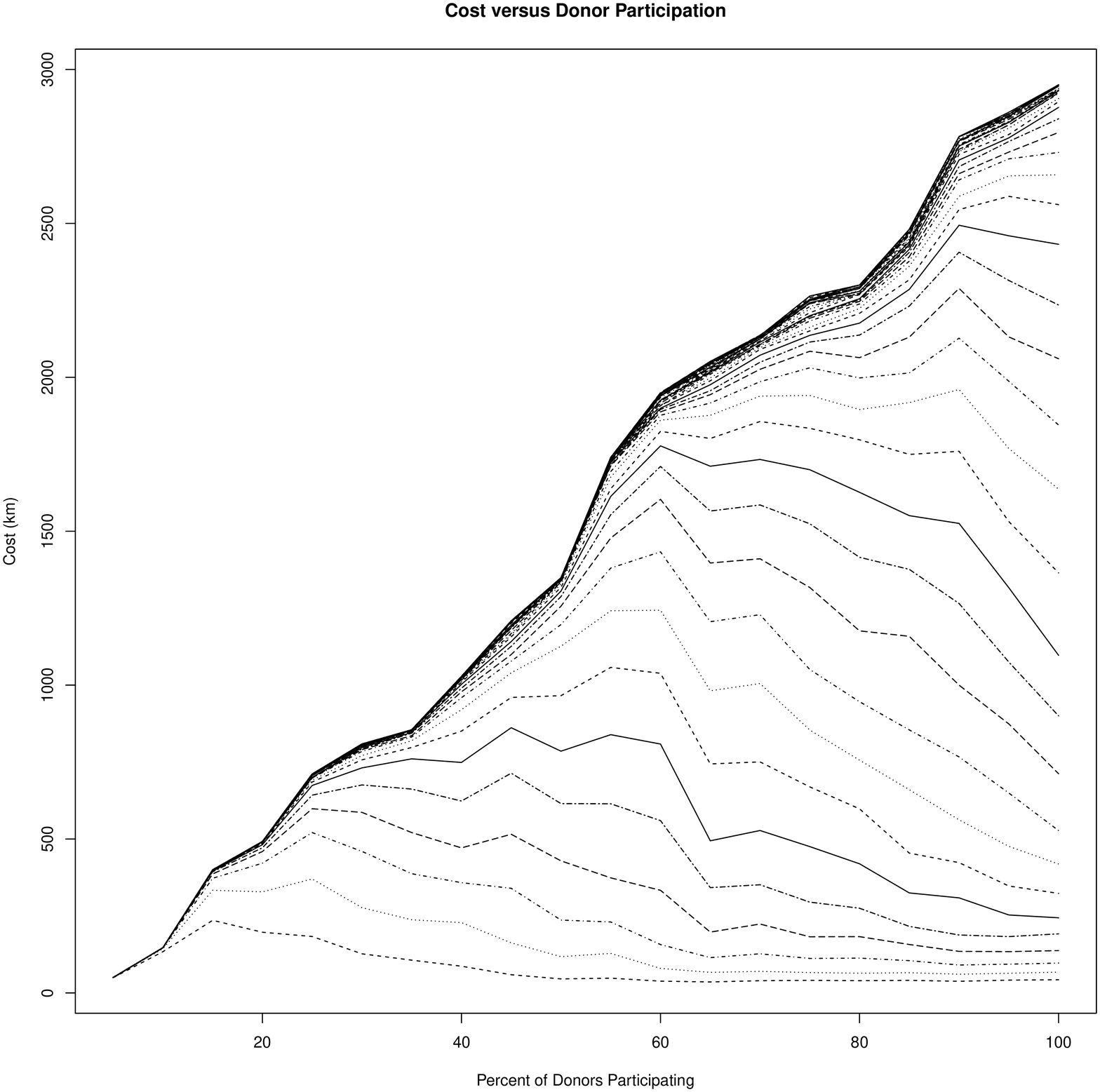}
\caption{Relationship between cost and percentage of participating donors in the complete supplier set. Each line corresponds to the cost curve associated with a different demand goal between 2,000 lbs (bottom-most line) and 48,500 lbs (line along main diagonal).\label{fig:costvpct}}
\end{centering}
\end{figure}

The other interesting thing to note about figure \ref{fig:costvrec} is that the shape of the curve appears to be linear for fewer suppliers, and trend towards exponential as the number of suppliers increases. We hypothesize that this is because the greater number of suppliers in the 156 case increases the densities of donors and allows for cost-saving optimizations when the demand goal is small. But, as the demand increases towards the saturation point, there are simply more donors that must be visited, quickly driving up cost. Figure \ref{fig:costvdem} helps clarify this relationship. In this figure, the cost is plotted as a function of the demand goal for each set of suppliers. We can see that up until the crossing point, the larger set of suppliers is able to satisfy the same demand with a smaller cost. These plots plateau at the point that they are no longer able to fulfill demand (and hence, have a horizontial asymptote at the cost of visiting all donors). Figure \ref{fig:costvpct} shows this relationship explicitly. To generate this graph, we take successively large random samples of the 156 supplier set and run a simulation for a fixed demand goal. In the plot, each line corresponds to the cost required for some fixed demand. In each line we can see roughly the same behavior: as the number of suppliers increases the cost goes up until a point is reached when the suppliers are able to meet demand (and hence we can optimize solutions to avoid some suppliers and drive down cost). After this point, which is different for each demand goal, the cost required decreases linearly (or superlinearly in some cases) as a function of the fraction of participating suppliers. This indicates that \textit{the cost of the food redistribution problem can be reduced simply by increasing the number of particpating donors}. 

\section{Conclusion}
\label{sec:fin}

In this paper we provide the first formal investigation of food recovery and redistribution problem as an optimization problem. To this end, we develope a novel model that can be used for Monte Carlo style simulation using fitted empirical parameters. For our experiments, we make use of data from a large food bank in northern central Colorado. We show that this model is able to reproduce the dynamics of the fundamental food redistribution problem faced by CFS. While we believe that this data is representative of a large class of similar regions with a mix of rural and urban environments, we are careful to remind the reader that our conclusions may not apply in disparite environments (i.e., dense urban or sparse rural). In future work, we hope to integrate additional data in order to broaden our conclusions and perform a more rigorous validation of our model.

Our chief experimental findings in this work are:

\begin{itemize}
\item Food supply (waste) events are heavy-tailed and can be well modeled with extreme value theory ``peaks over threshold'' models and the generalized Pareto distribution.
\item The efficacy of the food redistribution approach hinges on the ability to keep rescued food from perishing. Hence, refrigerated storage and timely transportation and processing are crucial to the success overall.
\item Despite the underlying heavy-tailed process and complexity of the model, the basic obtainable supply appears to be a linear function of the number of participating donors. Hence, doubling of the number of participating donors is likely to double the amount of food available.
\item The overall cost of food recovery can be reduced substantially simply by increasing the number of participating donors (and therefore creating more opportunity for food supply events to occur, when they are required by demand).
\item In the scenario we studied, using the data from CFS, we have shown that the food redistribution approach is a feasible method of subtantively reducing both waste and hunger. There is room for growth within the current paradigm, but it requires an increase in available funding.
\end{itemize}

In future work, we will expand this investigation to address additional questions. In particular we are interested in the question of whether this model can be scaled up to a state-wide or national level. It stands to reason that dense urban and sparse rural environments will produce substantially different cost and supply dynamics. However this is an open question. In 2008, there were approximately 85,200 grocery stores in the US \cite{USDL2011}, which works out to 0.3 chain supermarkets, 0.22 non-chain supermarkets, 3.04 grocery stores, and 1.8 convenience stores per zip code \cite{Powell2007}. In this work, we have not considered smaller potential donors, such as restaurants. One reason for this is that CFS is a Feeding America partner, which means that they are unable to accept food donations that are not in their original packaging \cite{tomreed}. Yet, this is a large potential food supply source.Iin \cite{Bloom2010}, Bloom suggests that the typical food waste associated with a restuarant is on average 3,000 lbs per employee, per year (or 123 lbs/day for a 15 employee restaurant). There were approximately 566,020 food service organizations in the US in 2007 \cite{USCensus2007}.
Clearly, there is no shortage of potential donors; the important question is whether they are well positioned for recovery and redistribution and whether the cost of rescue is acceptable.

An additional question is one of nutrition. In our current study, we looked at bulk lbs of food without concern for the type. Clearly, this is a large simplification that has bearing both on the economics of the problem (supply and demand) as well as the basic expiry of the
food. Currently, 88\% of grocery stores donate some dry goods, 51\% donate produce, and 31\% donate prepped food and meat \cite{Bloom2010}. Fresh and healthful foods are hardest for food banks to acquire since they have a limited shelf life (small $\epsilon$), which is negatively impacted by transportation and stocking time, and pickup limitations (how many pickups per week are possible). Optimizing pickup strategies, and sufficiently funding food rescue organizations so that they have the resources to pickup food when it is available might mitigate this problem. Feeding America's refrigerated trucks are a good start, but even these cannot captialize on smaller food waste events, which contribute substantively to the efficacy of the model as a whole. A complete solution may require rescue and redistribution at multiple scales and with varying technologies.

Although preliminary, our work here is an important first step towards understanding the dynamics and limitations of the food rescue and redistribution problem. By formulating the problem for optimization and studying it via simulation, we have been able to draw out fundamental
aspects of the underlying problem. In the end, we can present the positive result that this approach to food rescue and redistribution, despite its underlying complexity, can be considered a stable process where obtaining additional food is simply a function of having sufficient participating donors and funding to perform pickups. We hope that this work will help to spur interest in the area, equally among researchers who might be able to bring additional insight into the problem, businesses who can agree to donate their food waste, and policy makers who posess the ability to procure needed funding and resources for food rescue organizations to succeed.

\section*{Acknowledgements}

We would like to thank Community Food Share who generously provided us their data for analysis. We would also like to thank the numerous food rescue organizations and their tireless volunteers, whose work helped to inspire this research, as well as Professors Aaron Clauset and Elizabeth Bradley at the University of Colorado, Boulder, who provided valuable feedback on our initial investigation.

\bibliographystyle{abbrv}
\bibliography{food}

\end{document}